%% file: Gaskap_PASA_2012.tex
\documentclass[12pt,preprint]{aastex}
\usepackage{url}
\usepackage{epsfig}
\usepackage{xspace}
\def\kms{km~s$^{-1}$\xspace}
\newcommand {\scinot}[2] {#1\times10^{#2}\xspace}
\newcommand {\dd}       {{\mbox{$\rm \,deg^2$}}}
\newcommand {\hr}       {{\mbox{$\rm \,hr$}}}
\newcommand {\ddhr}     {{\mbox{$\rm \,deg^2\!/hr$}}}
\newcommand {\microns}  {{\mbox{$\, \mu {\rm m}$}}\xspace}

\newcommand {\hii}      {H~{\footnotesize II}\xspace}
\newcommand {\hi}       {H~{\footnotesize I}\xspace}
\newcommand {\hh}       {{\mbox{H$_2$}}\xspace}

\def\cm3{cm$^{-3}$\xspace}

\begin{document}

\input{gaskap_abstract_2012.tex}

\input{gaskap_sec1_2012.tex}

\input{gaskap_science_2012.tex}

\input{gaskap_technical_2012.tex}

\input{gaskap_refs_2012.tex}

\end{document}

%% file: gaskap_abstract_2012.tex
\title{\large GASKAP --- The Galactic ASKAP Survey}
\vspace{.4in}

\author{John M. Dickey}
\affil{University of Tasmania, Australia}
\email{john.dickey@utas.edu.au}

\author{Naomi McClure-Griffiths}
\affil{CSIRO Astrophysics and Space Science, Australia}

\author{Steven J. Gibson}
\affil{Western Kentucky University, USA}

\author{Jos\'e F. G\'omez}
\affil{Instituto de Astrof\'isica de Andaluc\'ia, CSIC, Spain}

\author{Hiroshi Imai}
\affil{Kagoshima University, Japan}

\author{Paul Jones}
\affil{University of New South Wales, Australia}

\author{Sne\v{z}ana Stanimirovi\'c}
\affil{University of Wisconsin, USA}

\author{Jacco Th. van Loon}
\affil{Keele University, UK}

\author{Andrew Walsh} 
\affil{James Cook University, Australia}

\author{A. Alberdi, G. Anglada, L. Uscanga}
\affil{Instituto de Astrof\'isica de Andaluc\'ia, CSIC, Spain}

\author{H. Arce}
\affil{Yale University, USA}

\author{M. Bailey}
\affil{Keele University, UK}

\author{A. Begum, B. Wakker}
\affil{University of Wisconsin, USA}

\author{N. Ben Bekhti, P. Kalberla, B. Winkel}
\affil{Universit\"{a}t Bonn, Germany}

\author{K. Bekki, B.-Q. For, L. Staveley-Smith, T. Westmeier}
\affil{University of Western Australia, Australia}

\author{M. Burton, M. Cunningham} 
\affil{University of New South Wales}

\author{J. Dawson, S. Ellingsen}
\affil{University of Tasmania, Australia}

\author{P. Diamond, J.A. Green, A.S. Hill, B. Koribalski, D. McConnell, J. Rathborne, M. Voronkov}
\affil{CSIRO Astronomy and Space Science, Australia}

\author{K. A. Douglas}
\affil{Dominion Radio Astrophysical Observatory, Canada}

\author{J. English}
\affil{University of Manitoba, Canada}

\author{H. Alyson Ford}
\affil{University of Michigan, USA}

\author{T. Foster}
\affil{Brandon University, Canada}

\author{Y. Gomez}
\affil{Universidad Nacional Aut\'onoma de M\'exico, Mexico}

\author{A. Green, J. Bland-Hawthorn}
\affil{University of Sydney, Australia}

\author{S. Gulyaev}
\affil{Auckland University of Technology, New Zealand}

\author{M. Hoare}
\affil{Leeds University, UK}

\author{G. Joncas}
\affil{Universit\'e de Laval, Canada}

\author{J-H. Kang}
\affil{Yonsei University, Korea}

\author{C. R. Kerton}
\affil{Iowa State University, USA}

\author{B-C. Koo}
\affil{Seoul National University, Korea}

\author{D. Leahy}
\affil{University of Calgary, Canada}

\author{N. Lo}
\affil{Universidad de Chile}

\author{F. J. Lockman}
\affil{National Radio Astronomy Observatory, USA}

\author{V. Migenes}
\affil{Brigham Young University, USA}

\author{J. Nakashima, Y. Zhang}
\affil{Hong Kong University, China}

\author{D. Nidever}
\affil{University of Virginia, USA}

\author{J.E.G. Peek\altaffilmark{1}}
\affil{Columbia University, USA}
\altaffiltext{1}{Hubble Fellow}

\author{D. Tafoya}
\affil{Kagoshima University, Japan}

\author{W. Tian, D. Wu}
\affil{National Astronomical Observatories, CAS, China}

\begin{abstract}
A survey of the Milky Way disk and the Magellanic System at the wavelengths of the 21-cm atomic hydgrogen (\hi) line and three 
18-cm lines
of the OH molecule will be carried out with the Australian Square Kilometre Array Pathfinder telescope.  The survey will study the distribution of
\hi emission and absorption with unprecedented angular and velocity resolution, as well as molecular line thermal emission, absorption, and maser 
lines.  The area to be covered includes the Galactic plane ($|b|<$ 10\arcdeg) at all declinations south of $\delta = +40$\arcdeg, spanning longitudes
167\arcdeg \ through 360\arcdeg \ 79\arcdeg \ at b=0\arcdeg, plus the entire area
of the Magellanic Stream and Clouds, a total of 13,020 square degrees.  The brightness temperature sensitivity will be very good, typically $\sigma_T \simeq $
1 K at resolution 30\arcsec \ and 1 \kms.  The survey has a wide spectrum of scientific goals, from studies of galaxy evolution to star formation, with
particular contributions to understanding stellar wind kinematics, the thermal phases
of the interstellar medium, the interaction between gas in the disk and halo, and the  
dynamical and thermal states of gas at various positions along the Magellanic Stream.
\end{abstract}

%% file: gaskap_sec1_2012.tex
\section{Introduction}

This paper describes a survey of the Milky Way (MW) Galactic plane, the Magellanic Clouds (MCs), and the
Magellanic Stream (MS) that will be carried out at $\lambda$ 21-cm and 18-cm to study the \hi
and OH lines.  The survey will reach an unprecedented combination of sensitivity and resolution,
using the revolutionary phased-array feed \citep[PAF,][]{Chippendale_etal_2010}
technology of the Australian Square Kilometre Array Pathfinder
telescope (ASKAP).  This telescope is currently under construction in Western Australia by the Australia
Telescope National Facility (ATNF), part of the Commonwealth Scientific and Industrial Research Organisation,
Astronomy and Space Science (CASS) branch.  The survey described here is among a group selected in 2009
to run during the first five years of operation of the telescope.  In the longer term, the design and
technology used in ASKAP may become the model for the ambitious Square Kilometre Array (SKA) instrument.
In the SKA era, surveys like the one described here will advance our knowledge of the Galaxy and its contents
in ways that will revolutionize astrophysics.  The project described in this paper is a step toward that
goal.

The Galactic ASKAP Survey (GASKAP) is the only approved ASKAP project that will have sufficiently high
velocity resolution to study the profiles of the \hi and OH lines in emission and absorption.
It is qualitatively different from the other planned ASKAP surveys in that high brightness
temperature sensitivity is the goal for much of the science, e.g. for detecting low column densities of gas.
This section describes the strengths of the ASKAP telescope for achieving this goal, and the reasons
behind the choice of observing parameters selected for GASKAP.  Section \ref{sec:science} discusses some of the 
scientific applications of the survey data, and the questions it will answer.  
Sections \ref{sec:simulations}, \ref{sec:data_products}, and \ref{sec:follow-up} describe the
planning for survey currently underway through simulations, specifications of the data products,
and follow-up observations.  

\subsection{The ASKAP Telescope}

The ASKAP telescope
\citep{Johnston_etal_2007,Johnston_etal_2008} is innovative in many ways, the most revolutionary being its focal plane
on which is mounted a phased-array feed and receiver array.  As currently designed and tested, the PAF uses
no feed horns or other concentrators of the radiation
focused by the 12m diameter primary reflector.  The radiation simply falls on the receiver array, which is carefully impedance matched to
minimize reflections and other losses, and contains 188 separate amplifier elements.  The signals are then further processed and combined
to make up to 36 independent beams with a total area on the sky of 30 square degrees (deg$^2$).  Each of these beams acts like the single-dish
primary beam of the interferometer, which is made up of 36 dishes and hence 630 baselines.  The wide field of view of
the small dish-plus-PAF combination leads to a very high survey speed.
With only the effective collecting area of a 72m diameter dish, ASKAP can observe a large area on the sky to a given flux density limit
faster than much larger radio telescopes that do not have PAFs.  

To detect an unresolved source, the critical telescope parameter is flux density sensitivity, $\sigma_F$, which is set merely
by the total effective collecting area, system temperature, bandwidth, and integration time
\citep{Johnston_Gray_2006}. 
But for a survey of extended emission that is distributed on angular sizes larger than the synthesized beam,
it is brightness temperature sensitivity, $\sigma_T$, that matters.  For a given total collecting area, the placement of the antennas of the array
determines the distribution of baseline lengths and hence both the maximum resolution and the brightness sensitivity.
The more widespread the antenna distribution the lower the filling factor, i.e. the 
covering factor in the aperture plane, $f$.  Lower filling factor results in worse brightness sensitivity, i.e. higher noise 
in brightness temperature, $\sigma_T$.

\begin{figure}[!b]
\begin{center}\epsfig{file=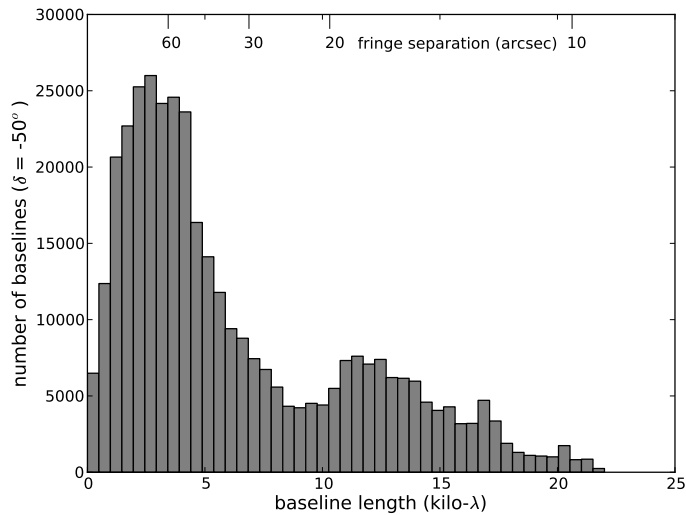,height=3in}\end{center}

\caption{\label{fig:baselines}  The ASKAP baseline distribution for a source at $\delta = -50$\arcdeg, from
\citet{Gupta_etal_2008}.  The two peaks at 2~--~6 k$\lambda$ (0.4~--~1.2 km) and 10~--~15 k$\lambda$ (2~--~3 km) are
designed to optimise the array for both extragalactic spectral line and continuum surveys.
For a Galactic survey, they are perfectly placed to measure \hi emission and absorption
as well as a combination of diffuse OH emission and OH maser emission. The y-axis gives the
number of one minute samples for a source at $\delta = -50$\arcdeg \ in a ten hour observation
at 1.42 GHz.}

\end{figure}

The ASKAP telescope is a general purpose instrument, with baselines up to 6 km in length, but most of the baselines
fall in two main groups, one with lengths between 400 and 1200 m, the other between 2 and 3 km (Figure \ref{fig:baselines}).
The ASKAP array was designed to provide optimum performance for extragalactic surveys of continuum and spectral line sources,
hence the two peaks in the baseline distribution.  Fortuitously, these two peaks are very well matched to the needs
of a Galactic \hi survey as well, with the dominant shorter baseline peak giving excellent brightness sensitivity at
beam sizes of 30\arcsec \ - 60\arcsec, while the longer baselines provide higher resolution
(10\arcsec) that will allow us to obtain sensitive \hi absorption spectra toward continuum
sources with flux densities as low as 20 mJy.  Similarly, emission from the 18 cm lines of OH occurs
both in very compact maser spots,
and in very widespread but faint thermal emission, while it appears in absorption toward compact, high brightness background continuum sources.  The 
ASKAP telescope design is an excellent match to the needs of a Galactic survey of emission and absorption in both the \hi and OH lines.

Following equations 3 and 4 in \citet{Johnston_etal_2007},
the ASKAP survey speed, $S(\sigma_T)$ in deg$^2$ h$^{-1}$ is a function of the rms noise in the brightness temperature, $\sigma_T$, as

\begin{equation}
 S(\sigma_T) \ = \ FoV \ B \ n_p \ \left( \frac{ \epsilon_c \ \sigma_T \ f}{T_{rec} \ \epsilon_s} \right) ^2 \end{equation}

\noindent
where the field of view, $FoV=30$ deg$^2$, the bandwidth $B$ is in Hz, the
number of polarizations $n_p=2$, the correlator efficiency
$\epsilon_c \le 1$,
the expected system temperature of the receivers
$T_{rec}= 50$ K,
the synthesis efficiency given by the taper, or weighting of the baselines in the mapping process 
$\epsilon_s \le 1$, 
and the effective aperture filling factor of the antennas
is $f$ where 

\begin{equation}
 f \ = \ \frac{ A \ \epsilon_A \ N \ \Omega \ \epsilon_s}{\lambda^2} \end{equation}

\noindent
with $\lambda$ the wavelength in m, $A=113$ m$^2$ the collecting area of a single antenna, $\epsilon_A \simeq 0.6$ the corresponding aperture efficiency,
$N=36$ the number of antennas, and $\Omega = 1.13\  \theta_s^2$\ the solid angle of the synthesized beam in steradians (sr) where $\theta_s$ is the full-width to half-maximum (FWHM) synthesized beamwidth.  
One of the options for ASKAP resolution has synthesized beamwidth,
$\theta_s=9.7 \times  10^{-5}$ rad = 20\arcsec \  (full-width to half-maximum), giving
$\Omega= 1.1 \times 10^{-8}$ sr, and for the $\lambda$21-cm line
$f / \epsilon_s \ = \ 5.8 \times  10^{-4}$. 
Using $B$ = 1 \kms $\simeq 5 \times 10^3$ Hz for the (heavily smoothed) effective bandwidth and $\epsilon_c = 
\epsilon_s = 1$ the ASKAP survey speed is:

\begin{equation}
 S(\sigma_T) \ = \ 0.10 \ \left(\frac{\sigma_T}{T_{rec}} \right)^2 \ \ {\textrm{\ deg}^2 \textrm{\  s}}^{-1} \quad \quad \textrm{(ASKAP)}\end{equation}

\noindent
and to get sensitivity $\sigma_T$ = 2.0 K 
this gives $S$ = 0.6 deg$^2$ h$^{-1}$
corresponding to integration time per pointing of $t_{int} = FoV / S$ = 50 hours.

For comparison, the Karl G. Jansky Very Large Array (VLA) 
C configuration gives resolution of about $\theta_s$=20\arcsec \ also, but
$N$=27, $T_{rec}\simeq 37$ K \citep[with  $\epsilon_A \simeq 0.6$]{Momjian_Perley_2011},
$A$=490 m$^2$ and $FoV$=0.32 deg$^2$, resulting in a filling
factor $f$/$\epsilon_s$~=~1.9$ \times 10^{-3}$ and a VLA C Array survey speed of

\begin{equation}
 S(\sigma_T) \ = \ 1.1\ 10^{-2} \ \left(\frac{\sigma_T}{T_{rec}} \right)^2 \ \ \textrm{deg}^2\ {\textrm s}^{-1} \quad \textrm{(VLA\ C\ array).} \end{equation}

\noindent
To get brightness temperature sensitivity $\sigma_T$ = 2 K requires $S$= 0.12 deg$^2$ h$^{-1}$, which is about one fifth the speed of ASKAP.

On the other hand, the Arecibo 305m telescope with the $N=7$ multibeam ALFA receiver
has $T_{rec}$ = 30K, $n_p$=2, $FoV = 3.8 \times  10^{-3}$ deg$^2$ 
for each beam
\citep{Goldsmith_2007}, and  

\begin{equation}
 S(\sigma_T) \ = \ FoV \ B \ n_p \ N \ \left( \frac{ \epsilon_c \ \sigma_T}{T_{rec}} \right)^2  \quad \quad \textrm{(Arecibo ALFA)}\end{equation}
\begin{equation}
 \quad \ = \ 270 \left( \frac{ \epsilon_c \ \sigma_T}{T_{rec}} \right)^2  \ \ {\textrm {deg}}^2 \ {\textrm s}^{-1} \end{equation}

\noindent
or $S$=170 deg$^2$ h$^{-1}$ for $\sigma_T$=0.4 K and $\epsilon_c=1$.  Note that the filling factor, $f=1$ for
a single-dish telescope with a filled aperture.  This survey speed is much faster than any aperture synthesis
telescope, but Arecibo's beam size of $\theta_s$=3.5\arcmin \ (at $\lambda$21-cm) is far from the 10\arcsec \ that ASKAP can achieve.

\subsection{Survey Description}

The GASKAP survey is one of ten approved survey science projects for ASKAP; its purpose is
to study the distribution of \hi and OH in the 
MW disk and the Magellanic System.  A summary of the survey
areas is presented in Table \ref{tab:areas}.
The GASKAP survey does not seek to cover the entire sky, as single dish surveys like GASS,
\citep{McClure-Griffiths_etal_2009}, LAB \citep{Kalberla_etal_2005,
Kalberla_etal_2010}, and GALFA-HI \citep{Peek_etal_2011a} have done. 
The interferometer sacrifices brightness sensitivity for resolution, so the GASKAP niche is to study regions
where the emission is bright, but with important structure on small angular scales and narrow velocity widths.
{\it \hi emission} from the Galactic plane and the MCs 
has brightness temperatures of tens to more than a hundred K,
so $\sigma_T$ of 2 K or less is sufficient to provide good signal to noise ratio.
{\it OH masers} will appear as
bright, unresolved spots of emission; for these the long baselines are needed to maximize the relative positional accuracy
at different radial velocities in the same source, thus allowing the precise determination of spatial-velocity structure. 
High flux density sensitivity is necessary for good astrometry so that OH maser positions can be compared with those of 
protostellar cores in star-formation regions, and with AGB/post-AGB stars from optical and infrared surveys.
For {\it \hi absorption} toward background continuum sources, resolution of 10\arcsec \ given by the longer
ASKAP baselines will allow the foreground emission to be subtracted accurately.  The
optical depth noise is then given by the strength of the continuum and by $\sigma_F$, not
by $\sigma_T$.  Thus the ASKAP telescope provides a great combination of high brightness temperature sensitivity
plus high angular resolution that matches the needs of {\it several different scientific applications}.
The scientific objectives of the GASKAP survey are discussed in more detail in section \ref{sec:science} below. 

To optimize a survey for mapping low surface brightness emission entails matching the telescope baseline distribution
to the scale of the structures of interest.  For a perfectly smooth brightness distribution, a single dish antenna is
the only tool to use, since interferometers have negative sidelobes that partially or completely 
cancel out the main beam response, depending on the
image restoration technique, e.g., clean or maximum entropy methods (Section \ref{sec:simulations}).  GASKAP will depend
on supplementary observations with single dish telescopes to fill in the emission with very large angle structure,
corresponding to very short baselines (``short spacing flux''), and ultimately any smooth background
(``zero spacing'', meaning a brightness constant over the whole sky).  The flux sensitivity of an interferometer
telescope is calculated by assuming that the source is unresolved, so that even the longest baselines do not suffer
any cancellation due to their finely-spaced positive and negative sidelobes.  The brightness sensitivity can
only be calculated given an angular size, using the baseline distribution of the antennas, as in 
Figure \ref{fig:baselines}.  Setting the specifications and strategy of a survey like GASKAP involves a process
similar to impedance matching, where the characteristics of the telescope are optimized for a particular range of
spatial frequencies, or angular scales of the distribution of the emission on the sky.
For ASKAP, there are relatively few baselines shorter than 100 m, so a single dish telescope of this
diameter or larger is optimum to fill in the short spacings.  The EBHIS \citep{Kerp_etal_2011} and GASS
\citep{Kalberla_etal_2010} surveys will be useful for this purpose.

Any particular interstellar structure, e.g., a shell, cloud, or chimney of whatever shape and size, has a 
corresponding flux distribution on the {\it u,v} plane, given by the Fourier transform of its brightness 
as a function of position on the sky.  The baselines of the telescope should sample this emission on the {\it u,v} 
plane as completely as possible, to give an image of the best possible fidelity, i.e. dynamic range.  For a survey,
the aggregate distribution of brightness over all angular scales throughout the survey region should be matched
by the baseline distribution and integration time of the telescope.  For the 21-cm line, the {\it u,v} distribution of
the brightness in the aggregate follows a power law function both in the Galactic plane at low latitudes 
\citep{Crovisier_Dickey_1983, Green_1993, Dickey_etal_2001},
in the MCs
\citep{Stanimirovic_Lazarian_2001},
and in the Magellanic Bridge 
\citep{Muller_etal_2004}.  
The power law is steep, having index -2.5 to -3.5 typically; so given the ASKAP baseline
distribution on Figure \ref{fig:baselines}, a reasonable goal for the survey is to obtain $\sigma_T$ somewhat below 2 K
on angular scales of 20\arcsec.  On smaller scales the emission is very faint, and the {\it u,v} coverage of the
telescope is relatively sparse for baselines longer than about 2 km, so very long integration time would be needed
to push beyond this sensitivity goal.  The longer baselines are useful for absorption spectra toward compact continuum background
sources, where brightness sensitivity is not the limiting factor.

Besides the MW and MCs, GASKAP will discover and map the dynamics of dwarf galaxies in the local volume
out to LSR velocities of $\pm$700 \kms.  Faint, irregular dwarfs
typically have narrow Gaussian \hi profiles due to their low rotational velocities.  Their linewidths are similar
to those of high velocity clouds (HVCs), but their velocity fields are generally very different.  Searching for gas rich, local group dwarf galaxies
is particularly important in the GASKAP survey area at low Galactic latitudes.  This science goal overlaps that of the
WALLABY project \citep{Koribalski_2012} that will cover the whole sky with velocity resolution of $\sim$4 \kms and
300 MHz bandwidth.

As a pathfinder to the SKA, the frequency range of ASKAP
was chosen to be 700 MHz to 1.8 GHz, with a maximum instantaneous bandwidth of 300 MHz.  This allows simultaneous
coverage of the 21-cm
line of atomic hydrogen (\hi) at 1420 MHz and three of the 18-cm OH lines at 1612, 1665, and 1667 MHz, but not 
the fourth at 1720 MHz.
For studies of \hi and OH in the MW, MCs, and MS, high velocity resolution is
critical.  The ASKAP spectrometer provides a total of 16,384 channels on each baseline.  These will be allocated to a few narrow
``zoom'' bands with fine velocity resolution.  For Galactic observations, a good
choice of
channel spacing is 976.6 Hz, giving a velocity step of 0.18 to 0.21 \kms (see Table \ref{tab:freq}). These narrow spectrometer channels 
will cover the four lines with LSR velocity ranges of $\pm 760$ \kms for \hi (7394 channels),
and $\pm 311$ \kms for the OH lines (3419 channels at
1612 MHz and 5571 channels covering both the 1665 and 1667 MHz lines together).  Since the OH main lines are separated by
just 352 \kms, blending of the two lines is possible in directions where the radial velocity of the emission spreads over
more than this amount.  This is not expected in the GASKAP survey area.
The allowed velocity range due to Galactic rotation in the inner
Galaxy, excepting the Galactic Centre, covers about $-$150 to +150 \kms maximum, depending on longitude.
In the MCs and MS the velocity ranges from about
+450 \kms in the leading arm \citep{Kilborn_etal_2000} to $-$400 \kms (LSR) near
the northern tip of the MS. 

\begin{figure}[!b]
\begin{center}\epsfig{file=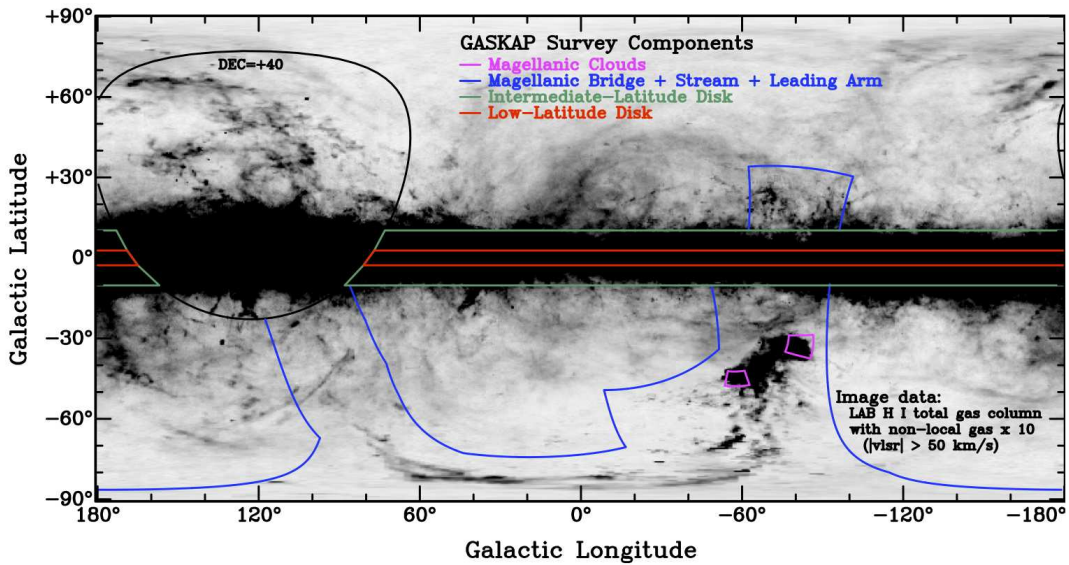,height=3in}\end{center}

\caption{ \label{fig:area}
 The GASKAP survey areas in Galactic coordinates, with \hi column densities from the LAB survey in the background.
The region north of $\delta$ = +40\arcdeg \ must be filled in from the Northern Hemisphere. The GAMES survey
described in section 5 will cover the region north of $\delta$=+40\arcdeg. }
\end{figure}

\subsection{Survey Parameters}

\begin{figure}[!b]
\hspace{-.4in} \epsfig{file=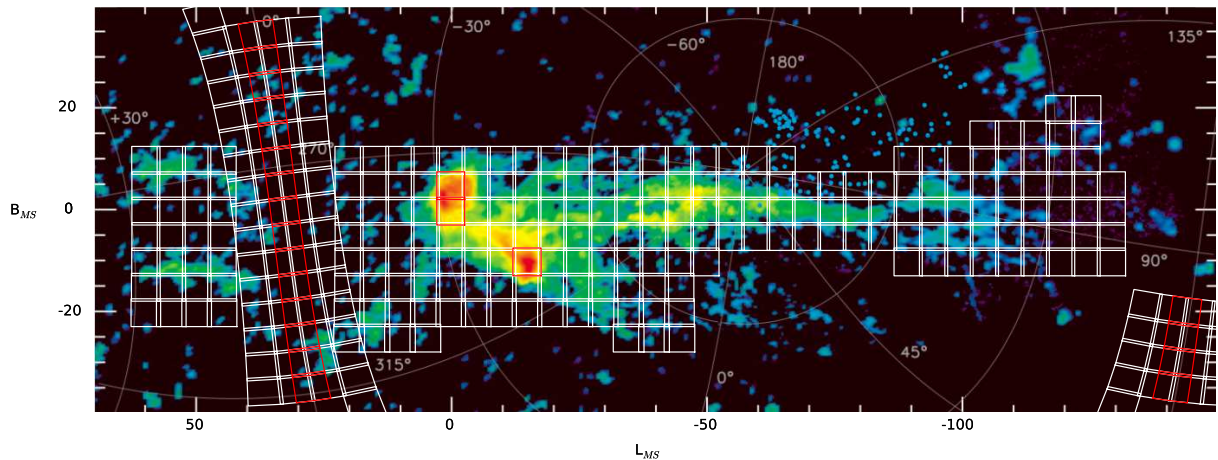,height=3in}

\caption{\label{fig:MSarea}
The GASKAP MS survey area with axes labelled in Magellanic Stream coordinates, and \hi column densities from the LAB survey in the background \citep{Nidever_etal_2010}. 
The white squares represent ASKAP pointings with the shorter integration time (12.5 hours), while the red squares are pointings that will be observed for either
50 or 200 hours.
}
\end{figure}

GASKAP will use three different survey speeds, with integration times of 12.5, 50, and 200 hours, that correspond to $S$ = 2.4,
0.6, and 0.15 deg$^2$ h$^{-1}$.  These translate to brightness temperature sensitivities at different angular resolutions as
given in Table \ref{tab:speeds}; smoothing to larger beam area gives much lower values of the noise in brightness tempearture,
$\sigma_T$.  The flux density sensitivity, given in the last column in Table \ref{tab:speeds},
 is only a weak function of angular resolution.  The lowest flux density noise level, $\sigma_F$,
is achieved with resolution 20\arcsec; at this resolution all baselines have roughly equal weights ($\epsilon_s = 1$).
Just three ASKAP fields are enough to cover most of the area of the MCs which will have 200 hour integration time per pointing
or ``Dwell Time'' (Table \ref{tab:speeds}, column 3).
A single line of 55 fields centred at $b=0$\arcdeg
\ each with 50 hours integration time, results in
a very sensitive survey
of the Galactic plane ($|b| < 2.5$\arcdeg) over longitudes 167\arcdeg \ through 360\arcdeg\ (the Galactic Centre) to 79\arcdeg.
An intermediate latitude strip of four rows of fields covers $|b| \leq 10$\arcdeg, centred at
$b = \pm2.5$\arcdeg \ and $\pm7.5$\arcdeg; these are observed at the fastest survey speed, with an integration time of
12.5 hours on each pointing.  
Finally, a wide area (about 6400 deg$^2$) of the MS is covered at the fastest survey speed (12.5 hours per pointing).  These areas are illustrated in Figures \ref{fig:area} and \ref{fig:MSarea}.

\begin{table}[]
\caption{ \bf Survey Areas \label{tab:areas}}
\begin{tabular}{|l|l|r|r|c|}
\hline
 & & {\bf Area} & {\bf Time} & {\bf Speed} \\
{\bf Component Name} & {\bf Location on Sky (see Figure \ref{fig:area})} & {\bf \dd} & {\bf \hr~} & $\!\!${\bf \ddhr}$\!\!$ \\
\hline
Low Latitude & $|b| < 2.5\arcdeg$, all $\ell$ for $\delta < +$$40\arcdeg$ & 1,650 & 2,750 & 0.60 \\
Intermediate Latitude & $2.5\arcdeg < |b| < 10\arcdeg$, all $\ell$ for $\delta < +$$40\arcdeg$ & 4,890 & 2,038 & 2.40 \\
Magellanic Clouds & LMC (2) + SMC (1) = 3 deep fields & 90 & 600 & 0.15 \\
Magellanic Bridge + Stream & $-135\arcdeg < {\ell_{ms}}^a < +66\arcdeg$, varying $b_{ms}$$^a$ & 6,390 & 2,662 & 2.40 \\
\hline
Total & & 13,020 & 8,050 & \\
\hline
\multicolumn{5}{l}{\footnotesize $^a$Magellanic Stream coordinates
\citep{Nidever_etal_2008}; see Figure  \ref{fig:MSarea}} \\
\end{tabular}
\end{table}

\begin{table}[!h] 
\caption{\bf Frequency and Velocity Coverage \label{tab:freq}}
\begin{tabular}{|lccc|}
\hline
 & Frequency & LSR Velocity & Channel\\  
Band & Range & Range & Spacing \\
& (MHz) & (\kms)  & (\kms) \\ \hline
\hi & 1416.795 -- 1424.016 & $\pm$762 & 0.206 \\
OH 1612 & 1610.561 -- 1613.900 & $\pm$310 & 0.182 \\
OH main lines & 1663.657 -- 1669.099 & $\pm$313 & 0.176 \\
\hline
\end{tabular}
\end{table}

\begin{table}[]  
\caption{\bf Survey Speeds and Sensitivity \label{tab:speeds}}
\begin{tabular}{|l|c|c|l|c|c|}
\hline
{\bf } & {\bf Map} & {\bf $\!$Dwell$\!$} & ~~~~~{{\bf $\sigma_T$} [K], $B = 5$~kHz} & {\bf $\sigma_F$} [mJy] \\
{\bf Survey Component} & {\bf Speed} & {\bf Time} & ~~~~~for $\theta_{FWHM} =$ & $B${\small =5 kHz} \\
{\bf } & {\bf $\!$\ddhr$\!$} & {h} & {\bf 20\arcsec ~~30\arcsec ~~~60\arcsec ~~90\arcsec ~180\arcsec} & \\
\hline
Magellanic Clouds & 0.15 & 200 & 1.01 ~0.48 ~0.21 ~0.12 ~0.05 & 0.5 \\
Low Latitude & 0.60 & 50.0 & 2.02 ~0.96 ~0.42 ~0.24 ~0.10 & 1.0 \\
Intermed. Lat.\ + MS\ & 2.40 & 12.5 & 4.05 ~1.91 ~0.85 ~0.49 ~0.20 & 2.0 \\
\hline
\end{tabular}
\end{table}

\section{Comparison to Other Surveys}
 
\begin{figure}[!b]
\begin{center}\epsfig{file=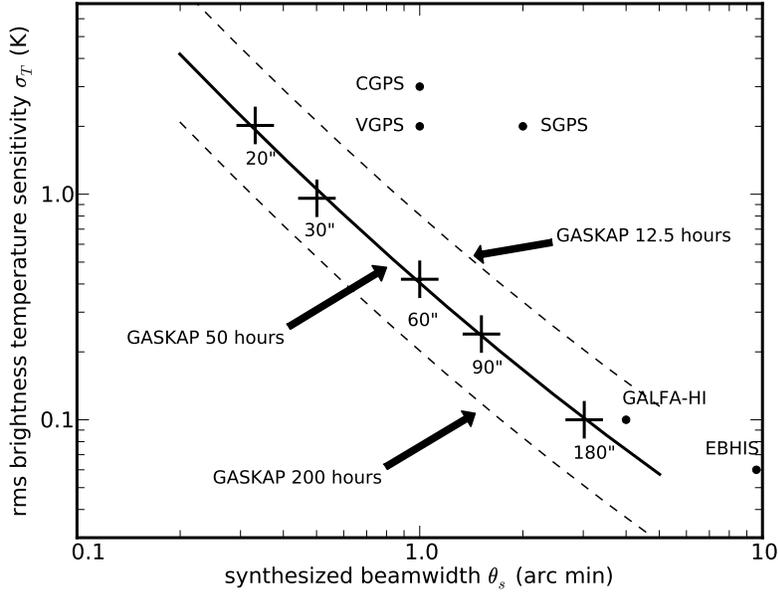,height=3.5in}\end{center}

\caption{\label{fig:Sig_B}  
The GASKAP brightness temperature sensitivity ($\sigma_T$) vs. resolution ($\theta_s$) 
with spectra smoothed to 1 \kms.
The solid curve represents the medium integration time of 50 hours per pointing,
while
the other two survey speeds have integration times four times longer or shorter, and hence they have sensitivities
a factor of two higher or lower, indicated by the dashed lines (see Table~\ref{tab:speeds}).
On the left ($\theta_s \la$ 20\arcsec) are combinations appropriate for
OH maser emission and 
\hi absorption at low latitudes, on the right ($\theta_s \ga$ 1\arcmin) are
combinations appropriate for low column density \hi in the Magellanic Stream and diffuse OH
emission in the Galactic plane.  \hi emission mapping at low latitudes will make use of
resolution from 20\arcsec \ to 1\arcmin, depending on the brightness and angular scales
of the emission in each field.  The GALFA-HI point is based on a 10s integration per beam area,
smoothed to resolution $\theta = 4$\arcmin.
}
\end{figure}

The brightness temperature sensitivity vs. angular resolution 
for the low latitude GASKAP survey component,
which will be observed using the intermediate mapping speed of 0.60
deg$^2$ h$^{-1}$ (middle row of Table \ref{tab:speeds}),
is illustrated in Figure \ref{fig:Sig_B},
along with the corresponding sensitivities of three recent
aperture synthesis surveys of the Galactic plane,
the Southern Galactic Plane Survey \citep[SGPS,][]{McClure-Griffiths_etal_2005}, the Canadian
Galactic Plane Survey \citep[CGPS,][]{Taylor_etal_2003}, and the VLA Galactic Plane Survey \citep[VGPS,][]{Stil_etal_2006}.  Also shown in Figure 
\ref{fig:Sig_B} are the sensitivities and resolutions 
of the Galactic ALFA
\citep[GALFA-HI,][]{Stanimirovic_etal_2008,Peek_etal_2011a}
seven-beam single dish survey and the 
lower resolution, all sky, 
EBHIS survey.
The GASS survey
would be off-scale in the lower right.  For GASKAP,
the lower angular resolution cubes (1.5\arcmin \ to 3\arcmin) are obtained
from the same data as the high resolution
images by smoothing in the image plane or by tapering more heavily in the $u,v$ (aperture) plane.
This taper reduces the effective collecting area of the array ($\epsilon_s < 1$)
for larger beamwidths, so that
the slope of the line in Figure \ref{fig:Sig_B} is not as steep as $-$2, as expected from equations
1 and 2 ($f \propto \sigma_T^{-1}$
with all other quantities fixed in equation 1, and $f \propto \Omega \propto {\theta_s}^{2}$ in equation 2,
but $\epsilon_s$ decreases weakly with increasing $\theta_s$).
The GASKAP survey is composed of different surveys done simultaneously; the output data from each one
will be useful for a variety of applications that
require different combinations of sensitivity and resolution, as
indicated in Figure \ref{fig:Sig_B} and discussed in the science topics section (\ref{sec:science}) below.

The tradeoff between resolution and brightness temperature sensitivity apparent in Table \ref{tab:speeds} is at once the limitation and
the great power of an aperture synthesis survey of diffuse emission.  For Galactic and Magellanic \hi emission,
the 30\arcsec \ and 0.2 \kms resolutions of GASKAP are a breakthrough, because they provide spectral line cubes 
comparable with the best images from surveys like those from space-based telescopes
in the far-IR (6\arcsec \ to 40\arcsec \ from the {\it Spitzer Space Telescope} at 24 to 160 $\mu$m)
and sub-millimetre (5\arcsec \ to 36\arcsec \ for the {\it Herschel Space Observatory} at 70 to 500
$\mu$m).
GASKAP data cubes will thus provide a well-matched comparison of \hi with ISM tracers at infrared, millimetre, and sub-mm
wavelengths.  With this survey it will finally be possible to obtain images of structures in the atomic (\hi) medium
with the richness and detail routinely available for the dust and molecular gas.

%% file: gaskap_science_2012.tex

\section{Scientific Goals of the Survey \label{sec:science}}

This section presents a series of short scientific discussions that
motivate the different applications of the GASKAP survey.  Because of the versatile capabilities of the telescope
and spectrometer, all of these goals are achieved simultaneously with the same observations.
For study of the interstellar \hi and OH of the Milky Way and Magellanic System, the fine velocity resolution of GASKAP
is critical, and even more so for OH masers.  As high spectral resolution is what distinguishes GASKAP from the other
ASKAP survey projects, the science goals mostly are formulated to take advantage of the narrow spectral channels and
high brightness sensitivity that will be obtained.

\subsection{ Galaxy evolution begins at home} 

One of the great challenges of
modern astrophysics is understanding how galaxies form and evolve.  This
is intimately connected with the outstanding problem of star formation:
as star formation transforms the interstellar medium (ISM), adding
heavy elements and kinetic energy, it determines the structure and
evolution of galaxies.  While modern cosmological theories can predict the distribution of dark matter in the Universe quite well, predicting the distribution of stars and gas in galaxies is still extremely difficult
\citep{Tonnesen_Bryan_2009,Tasker_etal_2008,Putman_etal_2009}.
The reason for this is the complex and dynamic ISM: simulations reach a bottleneck on scale sizes where detailed understanding of star formation, its feedback, and the interaction between galactic disks and halos need to be included
\citep{Stanimirovic_2010}.
To make advances in the area of galaxy formation and evolution, we must begin with our home neighborhood where the physics that drives this process can be exposed and studied in detail.

The GASKAP survey focuses on the generic physical processes that drive galaxy evolution by revealing their astrophysical basis
here at redshift of $z=0$.  GASKAP will provide a new and vastly improved picture of the distribution and dynamics of gas throughout the
disk and halo of both the MW and the MCs.  The data will provide the image
detail and broad range of scale sizes that are essential for a quantitative understanding of the physics of the gas in the MW and MCs, including the effects of radiation, shocks, magnetic fields, and the shapes of the gravitational potentials of the disk and halo.  Comparing the mixture of warm, cool, and molecular ISM phases in the MW and the MCs shows the variation of the heating and cooling rates with metallicity, and how these processes affect the star formation rate.  The MCs studied with GASKAP resolution in position and velocity will show two entire galactic systems in enough detail to trace the connection between star formation and gas infall and outflow. The specific astrophysical processes accessible to the survey are: the initial conditions for star formation and ISM phase transitions, the feedback processes in the ISM, and the exchange of matter between the disk and halo.
\vspace*{.1in}

\subsection{ Feedback processes --- wild cards in galaxy evolution}

Galaxy evolution is largely driven by star formation and the subsequent enrichment of the interstellar gas with heavy elements through red-giant winds and supernova explosions.  By undertaking an unbiased, flux-limited survey of OH masers in the MW and the MCs, GASKAP will image the gas at both ends of this cycle: the first stages of high-mass star formation and the last evolutionary phases of both the massive (8-25 M$_\odot$) supergiant progenitors of Type II supernovae and the more plentiful low- and intermediate-mass stars, i.e. oxygen-rich Asymptotic
Giant Branch (AGB) stars to Planetary Nebulae (PNe). 

The motion of the gas in the disk and halo traces both stochastic processes such as turbulence as well as discrete, evolving structures such as chimneys and shells.  We will study this motion primarily in \hi cubes that show the velocity structure of the diffuse medium, supplemented by more detailed maps of molecular clouds in diffuse OH emission and OH masers in regions of massive star formation.  These are the sources of the feedback that stirs up the gas \citep{Ford_etal_2008,Ford_etal_2010,Dawson_etal_2011a}.  
Clouds of neutral, atomic gas in the MW halo are excellent targets for the GASKAP survey because of its
high angular resolution.  Several examples have been studied with resolution of 30\arcsec \ with the VLA
\citep{Pidopryhora_etal_2009}; on this scale they show sharp density contrasts that suggest that they
are unstable in various ways, particularly to Rayleigh-Taylor fragmentation.  Thus cloud mapping
in \hi can reveal the evolution and dynamics of the gaseous halo.
GASKAP is designed to trace the effects of this feedback throughout the MW disk and lower halo.

\subsection{ How galaxies get their gas}

How much gas flows in and out of the disk through the halo, how fast does it flow, and what forces act on it along the way?  How do halo clouds survive their trip down to the disk?  These questions can be studied through \hi structure in the MS and HVCs \citep{Putman_etal_2011}, which reveals the conditions of the outer halo, and in the disk-halo interface, where the Galactic fountain constantly circulates \hi as evidenced by chimneys and \hi clouds \citep{Marasco_etal_2012,Lockman_2002,Ford_etal_2010,
Stanimirovic_etal_2006}.  
In low-mass galaxies, outflows are a determining factor in setting the rate of their gradual chemical enrichment.  
The 30\,Doradus mini-starburst in the LMC may be responsible for part of the MS, either as source or sink
\citep{Nidever_etal_2008,Olsen_etal_2011},
and the actively star-forming SMC has a porous ISM from which gas easily escapes, yet it is still extremely gas-rich (Figure \ref{fig:smc_sources}).
As outflow and accretion rates are expected to be a dramatic function of galaxy mass,
GASKAP's comparison of the disk-halo mass exchange in the MW and the MCs will probe the variability of
crucial physical parameters governing the large-scale gas flows
in three galaxies of very different masses in the range where these rates are expected to change dramatically.

Cosmological simulations predict that gas accretion onto galaxies is ongoing in the present epoch. The fresh gas is expected to provide fuel for star formation in galaxy disks \citep{Maller_Bullock_2004}.
Some of the \hi we see in the halo of the MW comes from satellite galaxies, some is former disk material that is raining back down as a galactic fountain, and some may be condensing from the hot halo gas \citep{Putman_etal_2009,Brooks_etal_2009}.  

Just how gas gets into galaxies remains a mystery.  The large number of gas clouds clearly visible in the Galactic halo is one potential source.  High velocity clouds fall into several distinct populations \citep{Wakker1991-III}, some associated with the Magellanic Stream and other clouds possibly decelerated by the hot halo \citep{Olano2008}.  One exemplar high velocity cloud system, known as the Smith Cloud \citep{Lockman2008}, is interesting as a rare example of a fast-moving massive stream ($\sim 10^7$M$_\odot$) close to the plane of the disk.  The Smith Cloud has roughly equal amounts of neutral and ionized hydrogen gas \citep{Bland-Hawthorn1998,Hill2009} and appears to have punched through the disk in the last 100 Myr \citep{Lockman2008}.  It is difficult to understand how this cloud has survived hitting the disk, or even its passage through the hot halo \citep[e.g.][]{Heitsch_Putman_2009}.   And yet, if we are to form a complete picture of the life-cycle of the Galaxy is imperative that we understand how systems like Smith's Cloud and the Magellanic System interact with the Galactic disk.   

GASKAP will provide complete, high resolution coverage of the Magellanic Stream and high velocity clouds associated with its Leading Arm as well as all high velocity clouds that come within 10 deg of the Galactic Plane.  A different ASKAP survey, WALLABY (B. Koribalski et al in prep.), will complement GASKAP by providing low spectral resolution images of Galactic, as well as extragalactic, \hi over the entire sky.  The all-sky coverage of WALLABY's Galactic \hi component will be useful for tracing full gas streams, like Smith's Cloud, over large areas of sky.  
By working with WALLABY to provide the context, GASKAP will be able to study the processes at work as high velocity cloud streams approach the Galactic disk.  The spectral resolution of GASKAP will allow measurement of the cooling, fragmentation and deceleration of \hi halo clouds as they near the plane and comparison with models such as those of \citet{Heitsch_Putman_2009}.


\subsection{ Disk-halo mass exchange and the energy flow in the ISM}

 For \hi emission, the GASKAP survey will provide the biggest improvement over existing survey data in the range $20\arcsec - 90\arcsec$ with a {\it tenfold} increase in resolution in most areas.  The
Galactic plane has been mostly covered at low latitudes ($|b|<1$\arcdeg \ or greater in some regions) by the combination of
the Canadian, Southern, and VLA Galactic Plane Surveys
\citep{Taylor_etal_2003,McClure-Griffiths_etal_2005,Stil_etal_2006}
with resolution ranging from 1\arcmin \ to 2\arcmin \ and brightness sensitivity 1.5 to 3 K rms, and GALFA-HI at
4\arcmin \ resolution and sensitivity 0.1 to 1 K rms \citep{Peek_etal_2007}. 
From these surveys we get a hint of the glorious images that GASKAP will produce.  The hierarchy of structure and motions of the ISM begins on scales of kiloparsecs, where we see how spiral arms influence gas streaming motions, shocks, and star formation
\citep{McClure-Griffiths_etal_2004, Strasser_etal_2007}. 
Continuing to $10-100$ pc scales we see shells, bubbles, and chimneys that trace the collective effects of many
supernova remnants and stellar winds
\citep{Normandeau_etal_1996,Stil_etal_2004,Kerton_etal_2006,McClure-Griffiths_etal_2006a,Kang_Koo_2007,Cichowolski_etal_2008}.
Moving down to scales of 1 parsec and smaller reveals tiny drips and cloudlets in shell wall instabilities
\citep{McClure-Griffiths_etal_2003,Dawson_etal_2011a,Dawson_etal_2011b}, small scale structure formed in colliding flows in the turbulent disk ISM
\citep{Vazquez-Semadeni_etal_2006,Hennebelle_Audit_2007} and ram-pressure interactions
between HVCs and the hot Galactic halo gas \citep{Peek_etal_2011a}.

Some recent \hi detections in dust shells around AGB and post-AGB stars mapped by ISO
\citep[e.g.][]{Libert_etal_2007} indicate that GASKAP could provide a larger catalog of such shells.
Considering that the distribution of PNe in height above the mid-plane, $Z$, follows that of the OH/IR stars, 
\citep[e.g. the MASH catalog][]{Miszalski_etal_2008},
many PNe will be found at intermediate $Z$ heights (0.3-1.0 kpc above the plane) and so
separated from most of the disk gas in the spectrum, and hence their shells may be detectable in \hi.
The interaction of stellar winds and supernova remnants with the 
ISM drives interstellar turbulence, seen in the ionized, neutral, and molecular phases with very similar spatial power spectra
\citep{Lazarian_Pogosyan_2000,Lazarian_Pogosyan_2006,Haverkorn_etal_2006}.
The GASKAP data will allow the power spectra of the ISM turbulence to be measured in a variety of different environments, with greater precision, and over a broader range of scales than any survey of neutral gas has done before.

\subsection{Phase changes in the gas on its way to star formation}

 What is the relationship between the atomic and molecular phases of the ISM in different interstellar environments?  GASKAP will trace these phases through \hi emission, \hi absorption, and diffuse OH emission.  Comparing them over a large area that contains many kinds of clouds, some forming stars and some not, will show how and where the gas makes the transition from one phase to another.  How does the temperature of the gas vary, and how are the different thermal phases mixed in different interstellar environments?  GASKAP will investigate this by comparing emission and absorption spectra to measure the excitation temperatures of the \hi and OH lines.  

 In \hi absorption, GASKAP will be an even greater advance over existing surveys than it is in \hi emission. 
We expect four extragalactic continuum sources per deg$^2$ with peak flux density, $F$, of 50 mJy or greater, and 10 sources per deg$^2$ with $F > 20$ mJy \citep{Condon_Mitchell_1984,Petrov_etal_2007}.  This is at least a factor of five more absorption spectra at this noise level than in any previous low latitude survey.
The rms noise is $\sigma_F \simeq$1 mJy in a spectral channel of width $\delta v$ = 1 \kms in the low latitude survey area. This gives optical depth noise of $\sigma_{\tau} \leq$0.02 in the absorption spectrum toward a source with $F$ = 50 mJy.  These spectra will have excellent signal to noise ratio in absorption.  The more abundant, fainter continuum sources will give absorption spectra of lower quality (e.g. $\sigma_{\tau}=$0.05 for $F$ = 20 mJy) that will be useful for statistical studies of the cool gas distribution, either by co-adding many spectra, or by integrating over velocity intervals much broader than 1 \kms.  

The big question that the absorption spectra will help answer is how the thermal interstellar pressure, which changes by several orders of magnitude from the
midplane to the lower halo and from the inner Galaxy to the outer disk, determines the mixture of warm and cool \hi 
\citep{Wolfire_etal_1995,Wolfire_etal_2003}.  Combining 21-cm absorption and emission spectra allows the excitation temperature
and column density at each velocity to be measured separately, giving a good estimate of the kinetic temperature.  The ambient pressure is tied to the equilibrium temperature through the cooling function of the \hi
\citep{Field_Goldsmith_Habing_1969,Dalgarno_McCray_1972}.
Separating the emission and absorption spectra requires good angular resolution to allow the emission to be measured very near to the continuum source.  The 
highest resolution of the ASKAP array, 10\arcsec, will be excellent for eliminating confusion due to small scale variations in the emission. Absorption line studies with GASKAP will be very valuable for understanding the huge range of physical conditions now being found in the neutral ISM \citep{Peek_etal_2011b,Jenkins_Tripp_2011}.
As an example, Figure 5 shows the level of improvement GASKAP will make relative to previous studies of absorption in the SMC. Only a handful
of HI absorption measurements exists for both the SMC and the LMC. White crosses in Figure 5 show the location
of radio continuum sources behind the SMC suitable for absorption measurements with GASKAP, while the white
circles show what has been done previously by \citet{Dickey_etal_2000}.


%
%
%
%

\begin{figure}[!b]

\begin{center}
\epsfig{file=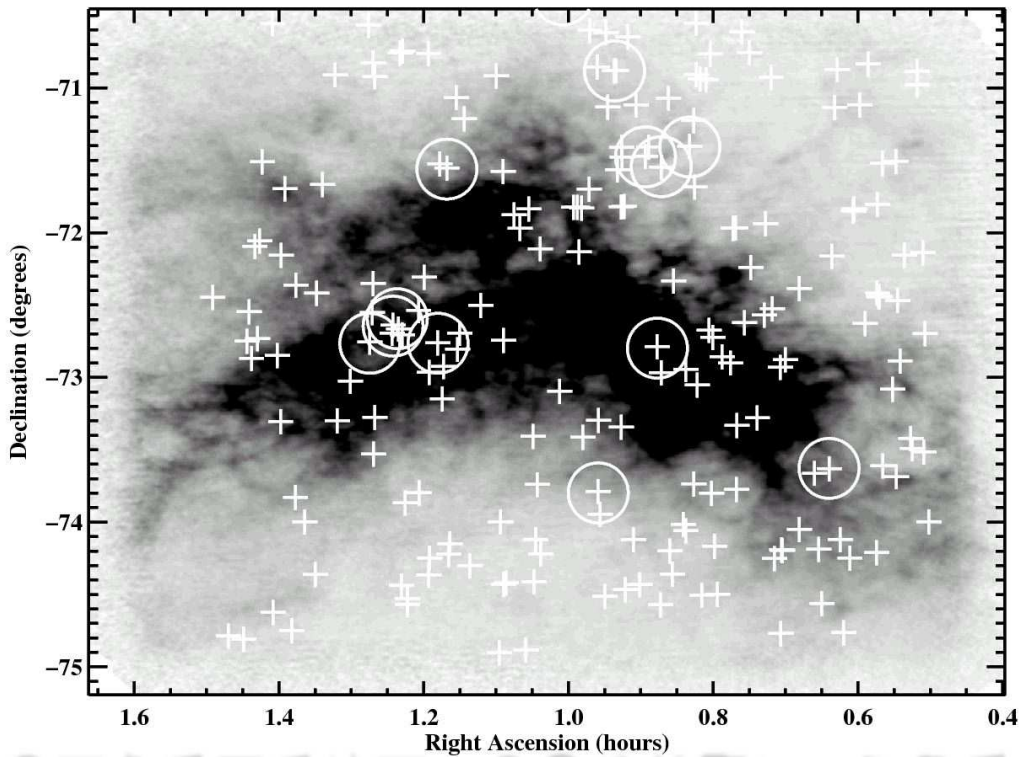,height=4in}
\end{center}

\caption{\label{fig:smc_sources} Locations of background continuum sources toward the SMC.
The circles show directions for which the \hi absorption spectra have already been measured.
The crosses show locations of sources bright enough to give good quality absorption spectra
with GASKAP.}
\end{figure}

The \hi\ absorption spectra toward continuum sources that GASKAP produces will yield a rich set of
gas temperature, column density, and velocity measurements over most of the
Galaxy.  Matched with these will be complementary, contiguous maps
of the cold \hi\ structure and distribution from
\hi\/ {\it self-absorption\/} (HISA) against Galactic \hi\ background emission
\citep{Gibson_etal_2000,Gibson_2010}.
HISA arises from \hh\ clouds as well as dense \hi\ clouds actively forming \hh, so it directly probes
molecular condensation prior to star formation
\citep{Kavars_etal_2005,Klaassen_etal_2005}.
A cloud's age can be estimated by
comparing its HISA and molecular content with theoretical models
\citep{Goldsmith_etal_2007,Krco_etal_2008}.  
GASKAP's low-latitude
survey will easily map the HISA from a $10-20$~K cloud with
$N_{HI} > 3 \times 10^{18}\, {\rm cm}^{-2}$.  This
sensitivity, enough to see the traces of \hi\ in molecular cloud cores,
plus cold atomic gas in \hi\ envelopes around the cores, will be applied to most of the
Galactic disk, enabling comprehensive population studies of \hh\/-forming
clouds, including their proximity to spiral shocks
\citep{Minter_etal_2001, Gibson_etal_2005}. 
GASKAP HISA will offer a rich new database for
rigorous tests of theoretical models of gas phase evolution in spiral arms
\citep{Dobbs_Bonnell_2007, Kim_etal_2008}
including phase lags between spiral shocks and star formation
\citep[e.g.][]{Tamburro_etal_2008}. 
On much smaller scales, the turbulent froth of HISA filaments that appear to be
pure cold \hi\ will be revealed at threefold finer angular and velocity
resolution in GASKAP than in all prior surveys, with sufficiently
improved sensitivity to follow their spatial power spectrum down to
sub-parsec scales.
This investigation will relate clouds' turbulent support to their stage of
molecular condensation.  Both 21-cm continuum absorption and self-absorption
toward Galactic objects,
including masers, are helpful for distance determinations \citep[e.g.][]{Anderson_Bania_2009,
Green_McClure-Griffiths_2011}.

\subsection{ Diffuse molecular clouds traced by extended OH Emission}

In addition to \hi\ spectroscopy, GASKAP will further enhance the
exploration of gas phase evolution with a new view of molecular
clouds.  The OH $\lambda$ 18~cm lines have long been used as an alternative to
the standard CO proxy for \hh, which is subject to the vagaries
of UV shielding, interstellar chemistry, and sub-thermal excitation
at densities below 10$^{3}$ \cm3 \citep{Liszt_Lucas_1996,Liszt_Lucas_1999,Grenier_etal_2005,Sheffer_etal_2008,Wolfire_etal_2010}.
However, diffuse OH emission is
typically about 100 times fainter than the \hi\ 21-cm line, and there
have been no large-area OH surveys since the ambitious survey of 
\citet{Turner_1979}
with the Green Bank 43m telescope.  ASKAP's new capabilities bring an unbiased and detailed
view of the OH sky within reach at last.  By simultaneously mapping cold
\hi\ self-absorption at 20\arcsec \ to 60\arcsec \ resolution and diffuse OH at 90\arcsec \  to
180\arcsec \ resolution (to achieve the necessary brightness sensitivity) GASKAP will
directly probe the \hh\ formation process by providing
a comprehensive \hi\/ + OH database of diffuse molecular clouds.
These data can be compared to 
quiescent evolutionary models
\citep[e.g.][]{Goldsmith_etal_2007,Liszt_2007}
and converging-flow dynamical models
\citep[e.g.][]{Bergin_etal_2004, Vazquez-Semadeni_etal_2007}
to address a key question in the field: {\it How long do \hh\ clouds take
to form from the diffuse ISM, and how does this affect star formation?\/}

Measurement of cloud total column density, temperature, mass, and other
properties will be needed to interpret mm-wave molecular line surveys,
infrared dust emission surveys and other future Galactic observations.  Of particular interest
would be a broad-based analysis of cold \hi, OH, CO, and dust in diffuse
clouds throughout the Galaxy to establish a common evolutionary clock for
clouds seen with multiple tracers.  GASKAP's $3\, \sigma$ OH 1667~MHz
sensitivity translates to a minimum detectable \hh\ column of
$\sim 1.0 \times 10^{21}\,{\rm cm}^{-2}$ for a 2~\kms wide line at
180\arcsec \ resolution, which is sufficient to sample the molecular content
of an $A_V \sim 0.6$ mag diffuse molecular cloud, or the early OH formation
in a dense molecular cloud.  This sensitivity will exceed that reached by 
\citet{Turner_1979} with
better velocity sampling and with angular resolution an order of magnitude sharper
over a larger and unbiased area, including the hitherto unexplored fourth
Galactic quadrant.  At the same time, OH absorption toward continuum sources
will be probed along with \hi absorption to show the excitation temperatures
of the 18-cm mainlines.  Assuming optical depths of a few times 10$^{-3}$ 
\citep{Liszt_Lucas_1996} and rms noise of 1 mJy for velocity resolution 1 \kms (Table \ref{tab:freq}), then
background sources brighter than about 1 Jy will show detectable absorption
in OH. Counting only extragalactic radio sources, there is one with flux density
greater than 1 Jy at 1.4 GHz on average 
every eighteen deg$^2$, but in the Galactic plane there will be many more \hii
regions and supernova remnants with flux densities well above 1 Jy, up to several
per degree of longitude for $| \ell | < 25$\arcdeg.  Thus we we expect
many OH absorption spectra for comparison with \hi absorption.

\subsection{OH masers in young and evolved stars}

\begin{figure}[!b]

\begin{center}
\epsfig{file=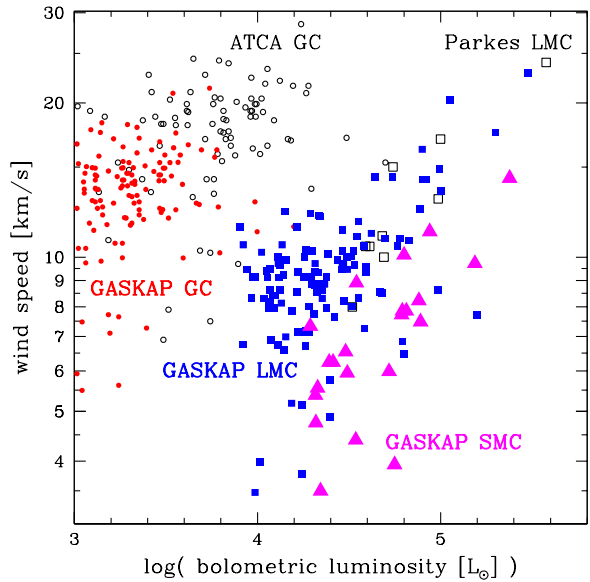,height=3in}
\end{center}

\caption{\label{fig:jacco_oh} Expected distributions of OH masers in
AGB stars and red supergiants based on empirical scaling relations in \citet{Marshall_etal_2004}.
Open symbols are known masers, with
circles for the Galactic Centre region and squares for the LMC; filled
symbols are predictions for GASKAP detections, with triangles for SMC masers.}
\end{figure}

OH masers allow us to study stellar birth and death, and give a
picture of Galactic structure and dynamics complementary to that shown
by the interstellar gas.
The OH portion of the GASKAP survey will allow study of the sites of high 
mass star formation and the old stars such as AGB stars, central stars of young PNe, and 
red supergiants that exhibit copious gas ejection in the last stages of their evolution. 
This phase of stellar evolution is a critical contributor to the chemical enrichment of the ISM.
OH maser spectra enable us to study the energetics of outflows from massive 
young stellar objects (YSOs) and the mass loss rates of dying stars by 
combining the measured radial velocities with estimates of the gas density in and around 
the OH emission region. 
With a sensitivity at least one order of magnitude better than previous 
spatially-limited surveys of OH masers \citep[e.g.][]{Sevenster_etal_2001,Caswell_Haynes_1987},
and taking into account the number of known OH maser sources to date 
\citep[$\sim$2300 in evolved stars,][]{Engels_etal_2010},
we expect to find several thousand new OH maser sources (Figure \ref{fig:jacco_oh}). Such a sensitive and unbiased 
survey will allow us to make statistical studies of the processes in these 
evolutionary phases as well as proper comparisons with results of Galactic surveys 
at other wavelengths. The total number of OH maser sources is also helpful 
to estimate the net mass ejection rate from stars in the Galaxy and to understand 
the circulation of processed material back into the interstellar medium. 

The accuracy of point source position measurement 
is $\sigma_{\theta} \simeq \frac{\theta}{\left(2\ \frac{S}{N}\right)}$, 
where $\theta = \theta_s$ is the FWHM of the synthesized beam, 
and $\frac{S}{N}$ is the signal-to-noise ratio of the source
\citep[e.g.][]{Reid_etal_1988}.
For strong maser emission ($>0.5$ Jy) with spatial-velocity structure, subarcsecond velocity gradients
will be measurable by finding the centroid of the emission in each velocity channel.
This will allow the tracing of 
kinematic structures such as disks or outflows in some OH maser stars, including highly collimated jets 
and ``super winds'' in some objects 
\citep{Sahai_etal_1999,Cohen_etal_2006}.
GASKAP will allow us to obtain a complete catalog of
evolved stars in special transition phases, of which there are only a few members
known so far. \citet{Zijlstra_etal_1989} catalog several examples of such evolved
stars showing both radio continuum and OH maser emission, which could be
extremely young PNe. Moreover, the shape of OH maser spectra can help identify
stars just leaving the AGB phase, when their spectra depart from the typical
double-peaked profile.  An extreme example are those evolved objects undergoing highly
collimated jets, such as ``water fountains'' \citep{Imai_etal_2007}.
These special types of objects are key to understanding the evolution and
morphology of PNe.

    Using the systemic velocities of masers in OH/IR stars, GASKAP will allow study of Galactic 
kinematics as traced by this stellar population with much more detail and spatial 
extent than previously \citep[e.g.][]{Baud_etal_1981,Sevenster_etal_1999}.
This will provide a picture of MW stellar dynamics that complements those from the gas 
and from the very different stellar populations that will be sampled by GAIA at optical wavelengths. 

Interstellar masers are one of the most readily detected signposts of locations where high-mass stars are forming. Surveys for masers are able to detect high-mass star formation regions throughout the Galaxy
\citep[e.g.][]{Green_etal_2009},
and the number of detected masers suggests that all high-mass star formation regions go through a phase where they have associated masers.  There are four different types of interstellar masers commonly observed in high-mass star formation regions -- ground state OH masers at 1665 and 1667 MHz, 22 GHz water masers, class II methanol masers (most commonly at 6.7 and 12.2 GHz) and class I methanol masers (most commonly at 36 and 44 GHz).  In addition to the common maser transitions, which are often (but not always) very strong, some sources also show maser emission from excited state OH transitions and/or higher frequency methanol transitions. These rarer masers are invariably weaker than the common masers observed in the same source
\citep[e.g.][]{Ellingsen_etal_2011}.
Studies of the pumping mechanism of the commonly observed star formation masers show that the observed transitions are less selective than some of the rarer maser transitions and are inverted over a wider range of physical conditions
\citep[e.g.][]{Cragg_etal_2002}.
Nevertheless, the detection of a particular maser transition towards a region indicates the presence of a relatively narrow range of physical conditions in the vicinity.  
\citet{Ellingsen_etal_2007}
suggest that the presence and absence of different maser transitions towards a region can be used as an evolutionary clock for the high-mass star formation process and proposed a time sequence involving all the common maser transitions.  Recent studies of methanol and water masers have further refined and quantified the maser-based timeline
\citep[e.g.][]{Breen_etal_2010a,Breen_etal_2010b,Ellingsen_etal_2011}.
The sensitive GASKAP study represents a unique opportunity to robustly test and refine the way that ground-state OH maser emission fits in the timeline.  

OH masers are generally thought to arise later than water and class II methanol masers in the evolution of high-mass star formation regions, but to persist after water and class II methanol switch off
\citep{Forster_Caswell_1989,Breen_etal_2010a}.
Most OH masers in high-mass star formation regions are strongest in the 1665 MHz transition, with weaker 1667 MHz emission and no 1612 or 1720 MHz maser emission.  There are however exceptions to this pattern
\citep[e.g.][]{Caswell_1999,Argon_etal_2003},
and GASKAP represents the best opportunity to determine if the currently known exceptions represent unusual objects, or perhaps short-lived evolutionary phases. 
\citet{Caswell_1999}
found that the majority of high-mass star formation regions with a 1612 MHz OH maser were not associated with class II methanol masers, suggesting that these are generally older, more evolved star formation regions. 
\citet{Argon_etal_2003}
found weak (200 mJy peak) 1665 and 1667 MHz OH masers toward the edges of a bipolar outflow traced by water masers towards the Turner-Welsh object near W3(OH).  The Turner-Welsh object is a protostar with a synchrotron jet which is too young, or not sufficiently massive, to have produced an \hii region. So the Argon et al. discovery appears to show some OH masers are associated with very different objects from the high mass proto-stars which traditionally show strong ground-state OH maser emission.  However, if the luminosity of the masers in the Turner-Welsh object is typical then no previous OH maser survey would have been sensitive to this class of object over a significant volume of the Galaxy.  With a 5-sigma detection limit in a 0.2 \kms spectral channel of approximately 10 mJy (for the low-latitude survey), it will be possible to determine what fraction of the water and class II methanol masers have associated OH masers with peak flux densities significantly less than 0.5 Jy and also to detect any weak OH masers which are not associated with other types of masers.  Simultaneous observations of three of the four ground-state OH maser transitions, combined with a growing amount of existing data on class I and II methanol masers and water masers, will enable us to substantially refine and improve the maser-based evolutionary clock for high-mass star formation regions.  The strong Zeeman splitting experienced by ground-state OH transitions means that many OH masers in star formation regions are highly (up to 100\%) circularly polarised and can be used to measure the total magnetic field in the maser region.  The spatial resolution of the GASKAP survey, combined with the (typically) complex 1665/1667 MHz spectra in high-mass star formation regions, means that in most cases follow-up observations, either of excited state OH maser transitions at higher frequencies or at higher spatial resolution, will be required to reliably determine the magnetic field from the star formation regions, although in some cases it may be possible to infer this directly from the GASKAP data products.

\subsection{Galactic metabolism in action in the Magellanic Clouds}

GASKAP will provide maps of both MCs with a 20\arcsec \ beam,
which is 3--5 times better than that of the seminal ATCA+Parkes \hi maps
of the LMC
\citep[1\arcmin \ beam]{Kim_etal_2003} and SMC \citep[98\arcsec \ beam]{Stanimirovic_etal_1999}, c.f. Figure \ref{fig:smc_sources}.
The GASKAP maps will have better velocity
resolution (0.2 \kms vs.\ 1.65 \kms for the earlier maps), and they
will have tenfold better sensitivity, $\sigma_T = 0.18$ K when smoothed
to 1.65 \kms and a 1\arcmin \ beam, compared to $\sim2$ K for the best existing
survey data.  The wide velocity range covered by the GASKAP maps will include
all high-speed gas. Crucially, the GASKAP maps will match the resolution
of the {\it Spitzer} SAGE survey maps at 70 \microns (18\arcsec) and the {\it Herschel}
HERITAGE maps at 160 \microns (12\arcsec). 
At the MC distance (50 to 60 kpc), resolution of $\sim$20\arcsec \ gives linear size
$\sim$5 pc, typical of supernova remnants and IR Dark Clouds.
The improved resolution allows direct links to be established between
the sources of stellar feedback and the ISM's response, as well as to
locate cold atomic clouds in absorption or 
self-absorption that are lost in the bright extended emission in lower
resolution data.

Two fundamental questions that GASKAP can answer about the gas in the
MCs are how effectively star formation can drive gas out of dwarf
irregular galaxies, and how differently star formation progresses in a
low-metallicity environment 
\citep{Krumholz_etal_2009}.  These are both critical issues for understanding the
formation of galaxies, through hierarchical merging of progressively smaller galactic building blocks.
Thus the GASKAP data on the MW and MCs will contribute to a fundamental astrophysical understanding
of the processes that dominate the epoch of galaxy formation.
Specific issues related to these questions include how much of the gas driven out of the MCs into the Bridge and Stream
will ultimately fall back to the MCs, how much will fall onto the MW disk, how much will blend into the hot halo, and
how much will be lost from the system altogether.
In its lowest resolution mode, GASKAP will
have the sensitivity to finally map the high-velocity gas that is seen
in absorption in front of much of the LMC
\citep{Lehner_etal_2009};
this will also finally allow the determination of a reliable
metallicity, and settle the question of whether this is a
chance-coincidence of a foreground Galactic cloud or whether it
represents an outflow from the LMC. 
GASKAP will provide hundreds of \hi absorption
spectra through the MCs; these will measure the mixture of warm and cool atomic gas
and the respective spatial distributions and kinematics of each phase.  The lower metallicity
of the ISM in the MCs will inhibit cooling in the medium.  This effect has
already been seen in absorption surveys that have been done with a
few background sources
\citep{Dickey_etal_1994,Marx_etal_1997},
but GASKAP will give a much higher density of lines of sight.

Simultaneously with the \hi observations, we will cover the MCs for the
first time with an unbiased, flux-limited survey in the OH lines.  Previous OH
observations were targeted and had worse sensitivity than GASKAP,
finding only the very brightest sources; the GASKAP OH survey of the
MCs will detect many more OH maser sources in star-forming regions.  Previous
OH maser observations of evolved stars in the MCs had rms sensitivities in excess of $\sim$10 mJy per 0.2 \kms\ channel. GASKAP achieves a ten times better sensitivity, yielding an expected two orders of magnitude more OH masers in the LMC
\citep[c.f.\ Fig.\ 12 in][]{Marshall_etal_2004} 
and the first such samples in the metal-poor SMC\@.  Large samples of masers in such low metallicity populations of cool giant and supergiant stars will test theories of the driving mechanism of the winds through measurement of their speeds from the double-peaked OH 1612 MHz maser profile \citep{Marshall_etal_2004,vanLoon_2006}. 
 \vspace*{.1in}

\subsection{ The Magellanic Stream -- a template for galaxy fueling processes}

Our closest example of a flow of gas from outside a galaxy making its way through the halo toward the disk is the MS.
GASKAP will survey about 5000 deg$^2$ of the MS, the Magellanic Bridge (MB), and the Leading Arm \citep[LA,][]{Putman_etal_2003}.
After smoothing the data cubes to 3\arcmin, we will achieve a 3-$\sigma$ sensitivity limit of N(\hi) $= \scinot{3.6}{18}$ cm$^{-2}$ per 20 \kms channel.  GASKAP's combination of angular resolution, sensitivity and spatial coverage is superior to all previous surveys of the Magellanic System
\citep[e.g.][]{Hulsbosch_Wakker_1988,Putman_etal_2003,Bruns_etal_2005,McClure-Griffiths_etal_2009}.  

Moving along the Stream from the MCs toward the Northern tip (near $\delta \sim$ +40\arcdeg), the \hi shows a wealth of small-scale structure down to the resolution limit of the existing surveys
\citep[3$'$ with the Arecibo radio telescope,][]{Stanimirovic_etal_2008}. 
It is not clear what drives the onset of this turbulent structure in the MS, or in accreting flows in general.  Various dynamical instabilities are expected to disrupt the MS
\citep{Bland-Hawthorn_etal_2007,Heitsch_Putman_2009}.  
Each has a distinct signature in the density and velocity fields, so that the GASKAP data will measure their relative importance.  Strong dynamical instabilities will lead to gas streamers and coherent structures, while thermal instabilities are expected to lead to fragmentation down to parsec and sub-parsec scales
\citep{Palotti_etal_2008,Heitsch_etal_2008,Burkert_Lin_2000}.   
H I clouds re-forming in the MW halo will have compact morphology and small velocity gradients, contrary to the freshly-stripped MS material that is expected to have a head-tail morphology.  These morphological and kinematic signatures are powerful diagnostics of the eroding agents essential for feeding the accreting material into the galaxy.  Such studies require high spatial and velocity resolution; they are not possible with existing survey data.

GASKAP observations will provide, for the first time, the angular resolution and sensitivity needed to resolve the interface regions between MS/LA/MB clouds and the surrounding hot gas; the physics of these interfaces is key to understanding how neutral gas flows both in and out through the halo. The temperatures and sizes of these interfaces determine the importance of thermal conduction for the dissipation of a ``fluff'' of small, neutral cloudlets.  Measuring the rate of this process is essential for understanding gas accretion in galaxies in general. The heart of this study will be a confluence of GASKAP observations and numerical simulations for cloud evolution by \citet{Heitsch_etal_2008} and \citet{Bland-Hawthorn_etal_2007}.

Recent studies
suggest that the MS is $\sim$40\% longer than previously thought, and it contains several long filaments
\citep{Westmeier_Koribalski_2008, Stanimirovic_etal_2008, Nidever_etal_2010}.
This extended filamentary structure is directly related to the history and evolution of the MS.  Numerical simulations that consider only gravitational interactions between the MCs and the MW 
\citep[e.g.][]{Connors_etal_2006} can reproduce such multiple filaments.
However, these simulations require two separate tidal encounters between the SMC, the LMC, and the MW
that are inconsistent with one of several recent proper motion measurements of the MCs \citep{Piatek_etal_2008} 
and complementary orbit calculations \citep{Bekki_2011,Diaz_Bekki_2011a,Besla_etal_2011}.
Using the combination of spatial coverage and angular resolution provided by GASKAP, we will study the
fine structure and kinematics, including the radial velocity gradients of individual MS filaments \citep{Stanimirovic_etal_2008}.
The results will discriminate among models of the orbital history of the MCs
and the role of both the luminous and dark-matter components of the MW halo \citep{Diaz_Bekki_2011b}.

\subsection{The cool neutral medium and star formation in low density environments} 

The MB presents a sliding scale of column density decreasing as one moves from the SMC toward the LMC\@.  Star formation is observed to happen
at the SMC end but not near the LMC \citep{Gordon_etal_2009,Harris_2007}.
Recent Arecibo and ATCA observations show the existence of a multi-phase medium in the MS \citep{Stanimirovic_etal_2008}.
\hi emission profiles show clear evidence for a warm and a cold component, at a distance of 60 kpc above the MW disk. 
\citet{Matthews_etal_2008}
detect an \hi absorption line in the MS, revealing a cool, neutral medium (CNM) core with a temperature of 70 K and \hi column density of $\scinot{2}{20}$ cm$^{-2}$.  Such a multi-phase medium with cold cores is unexpected in an environment such as the MS, based on the theoretical constraints on cooling/heating processes in the MW halo \citep{Wolfire_etal_1995}.  GASKAP will reveal and resolve many more cold cores in both the MS and the MB, allowing us to investigate the nature of the CNM in tidal tails, and thus the possible conditions for the formation of molecules, and ultimately stars, in low-density environments 
\citep{Heitsch_etal_2008,Bournaud_etal_2004,Schaye_2004}.

%% file: gaskap_technical_2012.tex
\section{Simulations, Imaging, and Algorithm Tests \label{sec:simulations}}

\subsection{\hi imaging}

Simulations and tests of the imaging pipeline are a significant part
of the GASKAP design study process.  GASKAP's imaging
requirements differ from the other ASKAP SSPs because \hi, and perhaps
OH, emission will fill the entire field of view on all angular scales
from arcseconds up to many degrees.  In addition to this large-scale
diffuse emission, we expect to see strong continuum absorption.  The
combination of strongly absorbed point sources in the midst of
large-scale diffuse emission presents an imaging challenge.  The
imaging pipeline is being tested both through simulations of the ASKAP
telescope's response to a modelled \hi sky and by using the ASKAP imaging
software, ASKAPSoft, on real interferometric data from the ATCA.

Two sets of simulations of the ASKAP telescope's response to a
modelled sky have been produced so far.  These simulations were performed by the CASS staff using
expected arrangements of the pointing centres for individual beams and the nominal beam shapes.
The GASKAP \hi simulations have been based on a sky model constructed by
scaling the pixel sizes from the GASS survey \citep{McClure-Griffiths_etal_2009,Kalberla_etal_2010}
to produce a model input cube
covering 8\arcdeg $\times$ 8\arcdeg \ with a native resolution of 10\arcsec.
The first round of simulations produced spectral line
cubes of 6\arcdeg $\times$ 6\arcdeg \ with resolutions of 30\arcsec,
60\arcsec, 90\arcsec, and 180\arcsec.  The simulations showed clearly
that deconvolution of the telescope beam (cleaning) is necessary to improve image quality at
all resolutions.  The simulations used multiscale clean 
\citep[MS-CLEAN,][]{Wakker_Schwarz_1988,Cornwell_2008,Rau_Cornwell_2011}
for deconvolution and this
appeared to work well.  For the \hi emission, combination with single
dish survey data is also necessary, since much of the overall sky
brightness is in very extended components that are filtered out by the
aperture synthesis process.  A second simulation has now been
produced, which includes \hi absorption towards continuum sources.
This simulation will be used not only to test the imaging pipeline but
to test our \hi absorption extraction pipeline.

Two clean algorithms have been tested, specifically MS-CLEAN and
maximum entropy.  Maximum entropy is a common 
algorithm for deconvolution of images with large-scale diffuse emission, like GASKAP 
\citep[see, e.g., ][]{Staveley-Smith_etal_2003,Stil_etal_2006,McClure-Griffiths_etal_2005}. 
Maximum entropy produces a
positive image with a compressed range of pixel values.  The
compressed range of pixels tends to produce a very smooth image and
the positivity means that negative sources, as produced by \hi
absorption where the continuum has been subtracted, are not
deconvolved.  MS-CLEAN is based on the traditional
\citet{Hogbom_1974}
version, but rather than working with point sources, the algorithm
works with multiple Gaussians of specified scale sizes.  Work by
\citet{Cornwell_2008,Rich_etal_2008} and \citet{Rau_Cornwell_2011} 
has shown that MS-CLEAN is much more effective at recovering large-scale
flux density than traditional H\"{o}gbom clean,
but it is untested in cases where emission fills the entire field of view.
MS-CLEAN is the default
deconvolution algorithm in the ASKAPSoft imaging pipeline.   

Working with existing \hi data from the SGPS Galactic Centre survey
\citep{McClure-Griffiths_etal_2012}, we have compared the results
of maximum entropy and MS-CLEAN.  These tests indicated that MS-CLEAN gave
qualitatively similar results to maximum entropy with the added
advantage that MS-CLEAN can deconvolve the absorbed continuum sources,
whereas maximum entropy cannot.  Furthermore the compressed pixel
range of maximum entropy means that the contrast between features in
the images is often better with MS-CLEAN than with maximum entropy.  

The GASKAP design study will determine the best method for combining
\hi single dish data with the ASKAP interferometric data.  A
discussion of various techniques for this combination is presented by
\citet{Stanimirovic_2002}.  This topic has become top priority now that MS-CLEAN has
been shown to be a viable deconvolution algorithm.  The goal is to determine the most
efficient and effective combination method, by 
exploring two options:  combination with single dish data during deconvolution,
by giving the single dish data as a model, and post-deconvolution
combination, by ``feathering'' the two images in the Fourier domain.



\subsection{Synthetic data cubes with OH maser emission}

One of the targets for GASKAP is the detection of Galactic OH masers,
in the transitions at 1612, 1665, and 1667 MHz. The satellite line at
1612 MHz
preferentially traces emission from evolved stars undergoing mass loss.
These are mostly stars on the AGB, but some
red supergiants, post-AGB stars, and
PNe are also OH emitters. The main-line transitions at
1665 and 1667 MHz are prominent in massive-star forming
regions, but they can also be found in evolved stars. OH maser emission
is spatially compact (typically less than $10^{14}$ cm) and spectrally
narrow (with intrinsic linewidths $\simeq 0.1$ to 1 km s$^{-1}$,
below thermal linewidths). Therefore, these masers are ideal targets for
source-finding algorithms.

Since GASKAP will detect more than $10^4$ maser sources, accurate and reliable source finding is crucial
for our science objectives.  To test the capabilities of source finding algorithms when dealing
with prospective GASKAP maser data, two simulated data sets of OH
sources have been generated to represent a single ASKAP field of view, one for star formation masers
($\simeq 200$ entries), and one for evolved stars ($\simeq 1200$ entries).

These two input  catalogs were processed using Miriad routines, to create
a FITS cube of the expected output from a single ASKAP field. The
final cubes had 1536 by 1536 spatial pixels
across a 6$^\circ$ $\times$ 6$^\circ$ field. The spectral resolution was 1.1 kHz
channels (equivalent to a velocity resolution of 0.2 \kms at the
1665-MHz frequency). There were 750 and 4000 spectral channels for
the cubes of star
formation and evolved objects, respectively.
Maser features  were assumed to be spatially unresolved and
to have Gaussian spectral profiles, with
widths ranging from 0.2 to 5 \kms (simulating different amounts of spectral
blending).
The beam size in the cubes is approximately 30\arcsec, with
10\arcsec \ pixels. They were created using robust weighting, applying a
Gaussian taper to the visibilities to reduce the weight of the long
ASKAP baselines. Arbitrary noise levels can be added to the simulated
data; for our first simulations, we have assumed that the
data came from a single
observing track of 10 hours of integration time per pointing.


\subsubsection{Simulated catalog of OH masers in evolved stars}

We have constructed
a simulated ASKAP observing field at 1612 MHz around the Galactic Centre.
This region is the most crowded one in terms of OH-emitting evolved sources
and therefore, it is used as the most challenging scenario to test the
efficiency of source finding algorithms and study issues such as confusion and
dynamic range limitations, considering that some OH masers can reach flux
densities of hundreds of Jy.

As of July 2010, the \citet{Engels_etal_2010} database listed
344 known stellar OH masers  at 1612 MHz within $3^\circ$
from the Galactic Centre. A histogram
of their flux densities is shown in Fig. \ref{fig:Jose-histo}a. However, this database
is obviously incomplete for weak masers, since no large-scale survey has ever
been conducted down to the sensitivities that can be reached with ASKAP.
To estimate the total number of sources we can expect to detect
in this field, we assumed that the actual flux density distribution of OH masers is
similar to that obtained by \citet{Sjouwerman_etal_1998}
in a smaller area
($37'\times37'$) around the Galactic Centre, after correcting by their
estimated completeness. We also assumed that the Engels database is
complete for the
strongest masers ($\log S_\nu(\textrm{mJy}) \geq 2.5$). Under these assumptions,
we obtained a total
estimate of $\simeq 669$ sources with $S_\nu \geq 3$ mJy (Fig. \ref{fig:Jose-histo}b) within $3^\circ$ from the Galactic Centre.
We note that the number of known stellar OH masers at 1612 MHz in this area is
only about half of our estimated number, which indicates the
usefulness of the sensitive and complete GASKAP survey of this type of sources.
To calculate the possible spatial, flux density, and velocity distribution of
the ``still undetected'' sources, we
considered some observational
properties of the known OH masers, such as their usual association with
SiO maser emission in AGB stars, and their mid-IR fluxes.

We considered as possible new sites of OH emission the known SiO masers
without detected OH emission.  The OH flux density was chosen
to be increasing with the MSX flux density at 21 $\mu$m, and conforming
to the flux density distribution in Fig. 1. The stellar velocities were directly
derived from the SiO spectra, and the OH masers where assumed to have a
double-peaked shape, i.e. modelled as two Gaussians, with a velocity separation of 15 km s$^{-1}$ (typical
of AGB stars). The known OH maser sources plus these prospective sources
selected from SiO maser sites amounted to a total of 556 sources.
To complete the catalog up to the expected 669 sources
in the field, we selected the 83 brightest MSX sources
fulfilling the mid-IR color
characteristics typical of AGB stars \citep{Sevenster_2002},
and without known OH and SiO masers.

\subsubsection{Simulated catalog of OH masers in star-forming sites}

\begin{figure}[!b]

\begin{center}
\epsfig{file=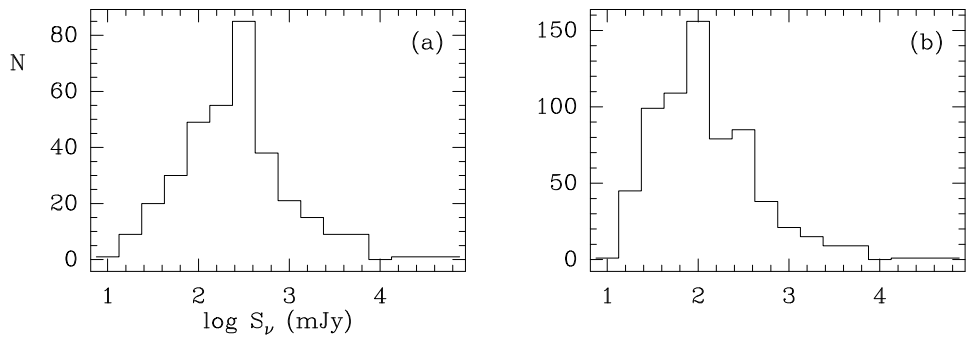,height=5in,angle=270}
\end{center}

\caption{ \label{fig:Jose-histo} OH Maser flux density distribution functions.
(a) The flux density distribution of 344
known OH masers within $3^\circ$ of the
Galactic Centre \citep[catalogued in][]{Engels_etal_2010}.
(b) The estimated flux
density distribution of the expected sources in the same region, after correcting
for completeness.}
\end{figure}

The star formation (1665 MHz) sample was derived from the 26 real sources
located within a $6^{\circ}$ diameter field centred on 338 degrees longitude
(a region of the Galactic plane with a high level of star formation
activity). It was supplemented with 15 extra (weak) sources based on
extrapolating the best current estimate of the star formation OH
luminosity function \citep{Caswell_Haynes_1987}
to the sensitivity limit of the GASKAP survey. The additional sources were located at the
positions of the brightest 6.7-GHz methanol masers with so far
undetected OH emission 
\citep[from the Methanol Multibeam Survey,][]{Green_etal_2009}, for which the OH masers are known to have a close
association with. In total there were 195 features across 41 sources
including 60 features across the 15 newly generated sources.  
These results are illustrated in Figure \ref{fig:Jose-histo}.

\section{Data Products \label{sec:data_products}}

The primary GASKAP data product will be high spectral resolution ($\sim$\,0.2 \kms)
cubes of the full survey region, covering \hi~emission over a velocity range
of $\pm$760 \kms, 1612\,MHz OH emission over $\pm$311 \kms and the two
OH maser lines at 1665 and 1667\,MHz over $\pm$311 \kms. In addition to this,
there will be two primary catalogues of spectra including OH maser emission and
\hi~absorption against background continuum sources. Finally, there will also
be high spatial resolution (2$^{\prime\prime}$ pixels, 10$^{\prime\prime}$ beam)
postage stamp cubes of all OH masers ($5^\prime \times 5^\prime$) and
\hi~and OH around all strong continuum sources
($60^{\prime\prime} \times 60^{\prime\prime}$), thereby including all
\hi~absorption features. Sensitivity maps and point-spread-function (PSF) maps
spaced at suitable points will be provided so that the sensitivity and PSF can be interpolated
to any point.

The data file size for a single image is anticipated to be 793\,GB. This assumes
that each field of view will be mapped out to dimensions of $7.5^\circ \times 7.5^\circ$, which will
result in substantial overlap of fields (separated by 5$^\circ$)
and provide a significant guard band around the edges of each PAF field of view.
On this basis, GASKAP will require approximately 1.0\,PB of
data storage space.


\section{Follow-up Surveys \label{sec:follow-up}}

A complete survey of the MW disk requires combination of observations by telescopes in the Northern and
Southern Hemispheres.  For GASKAP, the natural complement is a survey with the Westerbork Synthesis
Radio Telescope (WSRT) using the new array receiver Apertif 
\citep{Verheijen_etal_2008}.  The proposed
project, called the Galactic and Magellanic Emission Survey (GAMES), is intended to cover the remaining longitude
range of the MW (79\arcdeg $< \ell <$ 167\arcdeg) with the same latitude range and brightness sensitivity
as GASKAP.  The baseline distribution of the WSRT favours higher resolution, so GAMES may provide data
products with better positional accuracy and finer detail than GASKAP, but the primary data product
for GAMES
will be spectral line cubes that are tapered to match in spatial and spectral resolution 
as closely as possible with GASKAP.

High resolution follow-up on HI and OH continuum absorption and OH masers discovered by GASKAP 
will be possible using VLBI observations. ``Tiny structures'' of HI clumps as absorption in front of 
extended continuum sources will supplement our knowledge of
the power spectrum of the size distribution of HI clumps on scales down to 100 AU 
\citep{Davis_etal_1996,Deshpande_2000,Brogan_etal_2005}.
Similar absorption observations are possible for OH and H$_2$CO transitions at higher frequencies 
\citep[e.g.][]{Marscher_etal_1993}.
Similar VLBI observations of OH masers will reveal the details of the 
spatial and kinematic structure of gas around young stellar objects and dying stars. VLBI astrometry of OH masers 
will provide an opportunity to study the dynamics of the MW as a whole. In contrast to recent
maser astrometric observations that concentrate on star forming regions in the Galactic thin disk 
\citep[e.g.][]{Reid_etal_2009},
parallax and proper motion observations of stellar OH masers may extend the 
exploration of MW dynamics to the Galactic thick disk.  In the Magellanic Clouds, maser
observations using VLBI astrometric techniques hold the promise of showing the proper motion and someday even
the parallax of the Clouds.
Additionally, full Stokes polarimetric observations of the OH masers with the Australia Telescope Compact Array, such
as those being conducted for the MAGMO project \citep{Green_2010}, will enable exploration of the properties of the in
situ magnetic field. 

\begin{figure}[!b]

\begin{center}
\hspace{-.42in} \epsfig{file=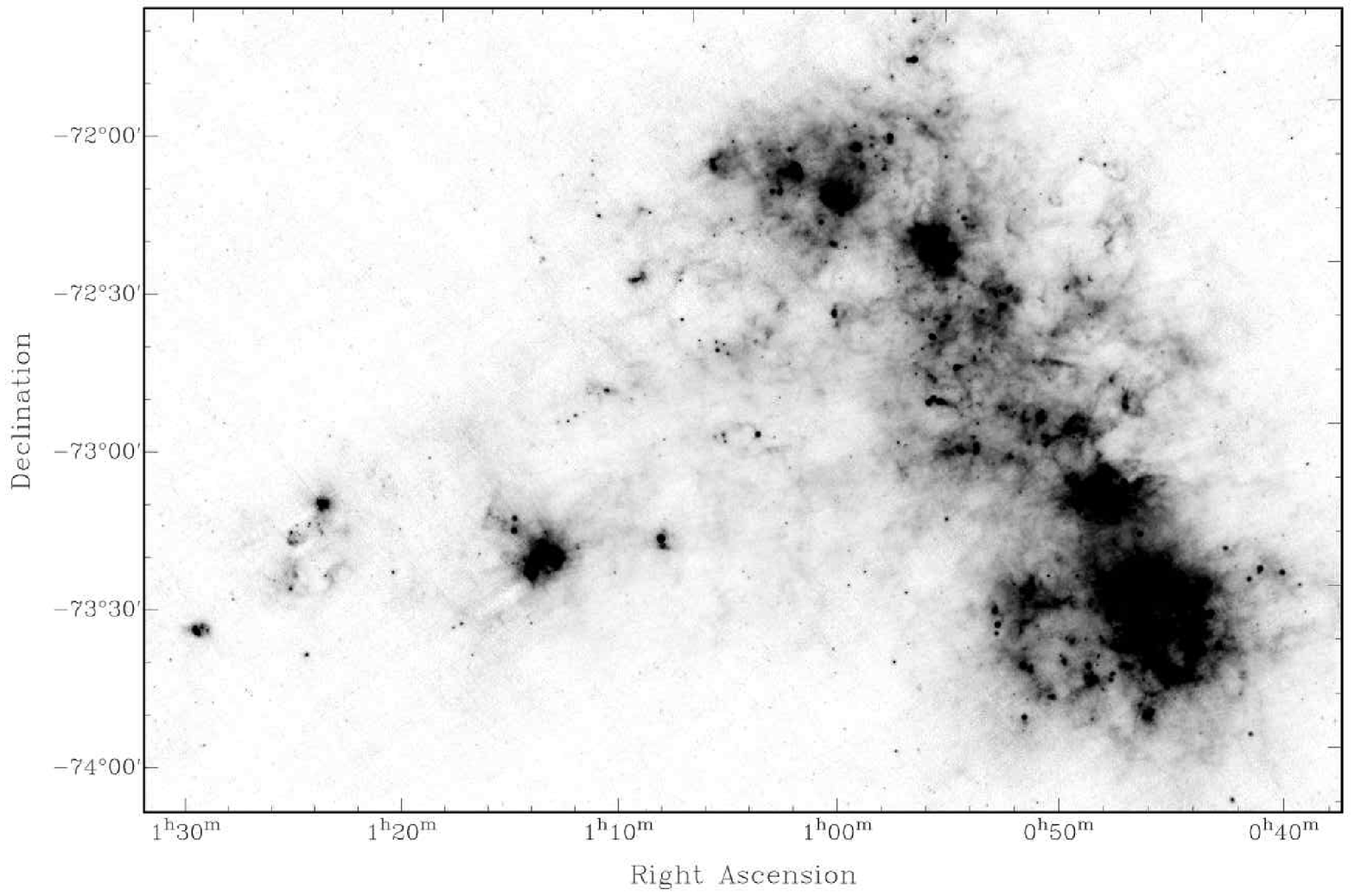,height=4.9in,angle=270}

\epsfig{file=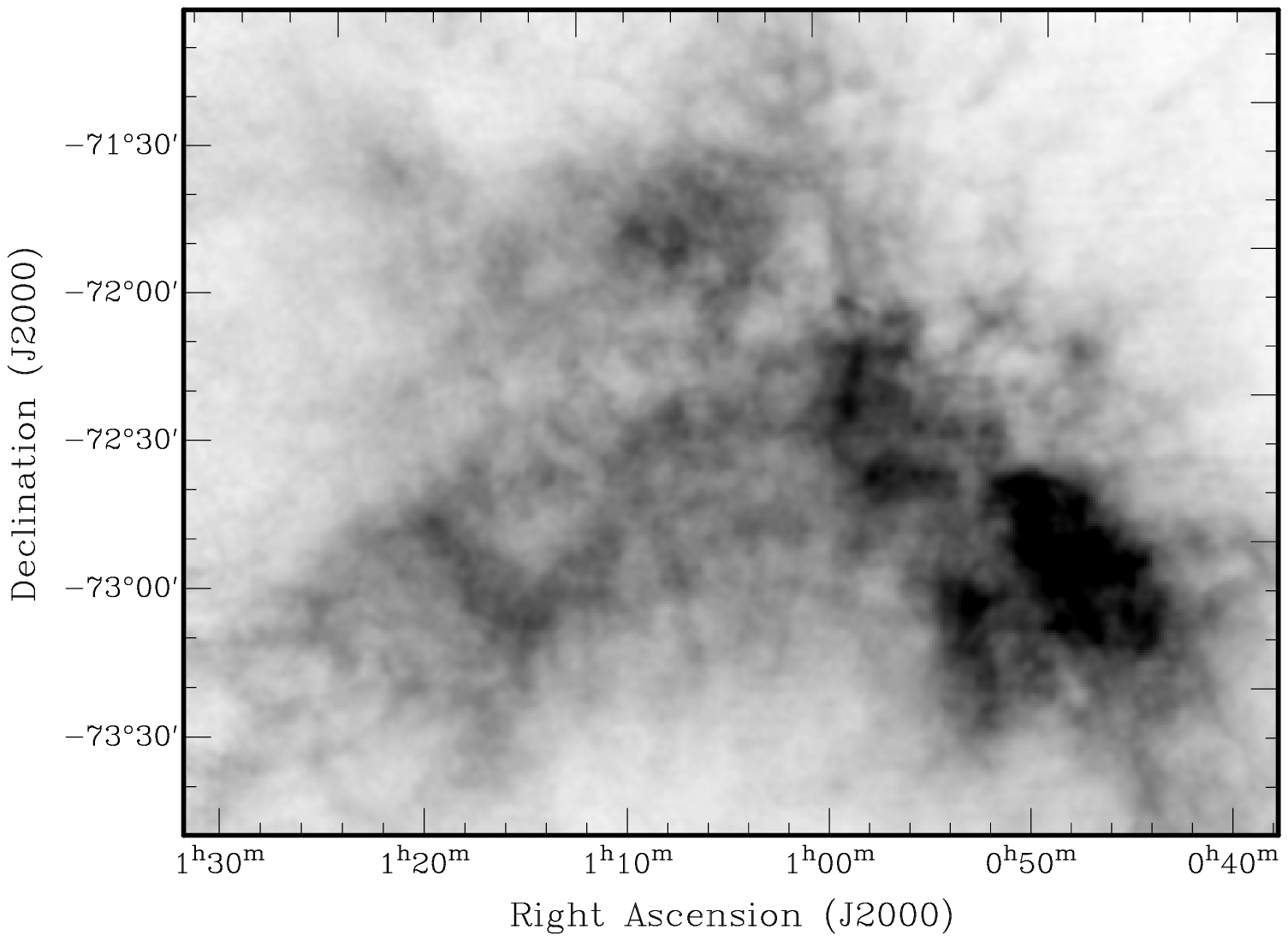,height=4in,angle=270}
\end{center}

\caption{ \label{fig:jacco_smc} Two images of the SMC.  On top is the recent SAGE-SMC
{\it Spitzer} Legacy image at 70 $\mu$m
\citep{Gordon_etal_2011}
tracing dust emission in the far-IR with resolution 18\arcsec.
Below is the best existing image in the 21-cm line, tracing total HI column density with
resolution 1\arcmin \ 
\citep{Stanimirovic_etal_1999}.  The GASKAP survey will improve the resolution
in the 21-cm line by a factor of three, nearly matching the resolution of the {\it Spitzer} image.}

\end{figure}

The GASKAP survey will be the most compelling scientific advertisement
for the Square Kilometre Array because of its image quality. 
As an example of the value of comparing surveys of the interstellar medium using different
tracers, and also as an illustration of the power of imagery with high spatial resolution and
dynamic range, Figure \ref{fig:jacco_smc} shows two views of the SMC.  On the top is an
image of the dust made with data 
from {\it Spitzer} at 70 micron wavelength (Gordon et al.\ 2011),
and below
is the best current image of the 21-cm column density of HI \citep{Stanimirovic_etal_1999}.  
Far-infrared dust maps and atomic hydrogen maps can be combined to estimate the amount and
distribution of {\it molecular} gas, which is most directly associated with the star formation process
\citep[e.g.][]{Bolatto_etal_2011},
but for this to be accurate these maps need to have
comparable angular resolution.
The results of previous Galactic surveys
demonstrate that GASKAP will produce images of structures in the ISM
with stunning detail and compelling aesthetic appeal.  For the general
public these images may be the most appealing results to come
from the entire ASKAP effort.  The GASKAP survey is designed to have
the maximum possible impact to further the Square Kilometre Array
project. The results will appeal to astronomers and non-astronomers
and contribute to fields of study in a broad base of theoretical and
observational research well beyond the traditional radio astronomy community.

\section{Summary \label{sec:summary}}

The GASKAP survey is the only approved survey science project for the ASKAP telescope that
will concentrate on spectroscopy of the Milky Way and Magellanic System with narrow velocity
channels covering the \hi and OH lines at wavelengths 21 and 18 cm.  Because it is the only
such survey, GASKAP necessarily represents a combination of many different scientific applications and
objectives.  The results of the survey will be valuable for questions in cosmology, 
galaxy mergers and accretion, the structure and dynamics of the Magellanic Clouds and Stream and
of the outer halo of the Milky Way through which they move.  The survey will discover and catalog
thousands of OH masers, tracing outflows from evolved stars, as well as star formation
regions and high-mass protostars.  The Milky Way interstellar medium will be studied through 
different tracers that show different thermal phases; one of the most difficult to study in any
other way is the cool, atomic medium that will appear in absorption toward background continuum
sources and sometimes toward the \hi emission itself.  The survey team includes specialists in all
these topics and more, but the data will be made public as soon as its quality is assured.
There is no proprietary period for the results of the survey, and archiving and distribution
of preliminary survey data will begin long before the observations are finished.

The ASKAP project timetable depends on technical and administrative issues.  The current plan as of
early 2012 calls for test observations to be made with a small subset of the antennas, the
Boolardy Engineering Test Array (BETA), in later 2012, with some 12 antennas ready for observing
by mid-2013.  The hope is to begin taking test data for the GASKAP survey soon after that, with
full array observations underway by 2014.  Recent single dish and interferometer
tests of the PAF on two 12m dishes have been
very encouraging, and the BETA array will soon be making aperture synthesis maps of test
fields in various environments, including Galactic plane \hi.  Meanwhile computer simulations of the telescope response
and resulting spectral line cubes are under intense study by members of the GASKAP team
for planning purposes, particularly in source finding (continuum and OH masers), survey strategy, 
beam deconvolution, and data quality assurance.